\documentclass[onefignum,onetabnum]{siamart250211}

\usepackage{graphicx} 
\usepackage[normalem]{ulem}
\usepackage{color}
\usepackage{amsmath,amssymb,amsfonts}
\usepackage[percent]{overpic}

\newcommand{\R}{\mathbb{R}}
\DeclareMathOperator*{\E}{\mathbb{E}}

\newtheorem{Def}[theorem]{Definition}

\usepackage{epstopdf}
\usepackage{algorithmic}
\ifpdf
  \DeclareGraphicsExtensions{.eps,.pdf,.png,.jpg}
\else
  \DeclareGraphicsExtensions{.eps}
\fi

\usepackage{multirow}
\usepackage{colortbl}
\usepackage{subcaption}

\newsiamremark{remark}{Remark}
\newsiamremark{hypothesis}{Hypothesis}
\crefname{hypothesis}{Hypothesis}{Hypotheses}
\newsiamthm{claim}{Claim}

\newcommand{\jr}[1]{{\color{red}{ #1}}}

\title{Dynamic threshold curves and response precision in forced excitable systems\thanks{Submitted to the editors DATE.
\funding{Jonathan E. Rubin acknowledges support from the National Science Foundation (NSF) under Award DMS-1951095. Justyna Signerska-Rynkowska acknowledges support from National Science Centre (Poland) under grant 2019/35/D/ST1/02253 and the support of Dioscuri program initiated by the Max Planck Society, jointly managed with the National Science Centre (Poland), and mutually funded by the Polish Ministry
of Science and Higher Education and the German Federal Ministry of Education and Research.}}}

\author{Jonathan E. Rubin\thanks{University of Pittsburgh, Pittsburgh, PA 15260 USA  (\email{jonrubin@pitt.edu}).} \and Justyna Signerska-Rynkowska\thanks{Institute of Applied Mathematics, Gda\'nsk University of Technology, ul. Narutowicza 11/12, 80-233 Gda\'nsk, Poland (\email{justyna.signerska@pg.edu.pl}).} \and Jonathan Touboul\thanks{Department of Mathematics and Volen National Center for Complex Systems, Brandeis University, Waltam MA 02453 (jtouboul@brandeis.edu).}}

\headers{Dynamic thresholds and response precision}{J.E. Rubin, J. Signerska-Rynkowska, J. Touboul}

\begin{document}

\maketitle

\begin{abstract}
We investigate here various properties of the responses of excitable systems subject to periodic forcing and noise. While the properties of intrinsic oscillators, subject to added periodic signals, are well understood, much less is known about the factors that determine the \emph{response precision} of excitable units, intrinsically at rest, when activated by periodic forcing and stochastic noise.  One motivation for considering this issue comes from the behavior of auditory neurons. These neurons reportedly have the ability to fire spikes in a precise range of phases in response to incoming sound waves, a behavior for which the mechanism is unknown. To account for such a response precision, we introduce the notion of dynamic threshold curve (DTC), which estimates at each time the effective likelihood that noise will subsequently generate a spike.  The DTC effectively summarizes, in a single curve, a representation of the response precision of an excitable model, as we demonstrate by showing that the distribution of spike times produced in this setting is well captured by the first passage time of a simple, Gaussian stochastic process to the distance to the DTC. This result shows that peaks and troughs of the DTC, but also their slopes, convey fine information about spike timing in response to noise. In particular, it explains properties of Type 2 and Type 3 excitable cells studied previously and provides a framework to predict the DTC properties necessary to support the response precision of auditory neurons, as we illustrate in a well-established auditory neuron model.

\end{abstract}
\begin{keywords}
phase-locking, periodic forcing, auditory neuron, stochastic process, type 3 dynamics
\end{keywords}

\begin{MSCcodes} Primary: 37C60, 37N25; Secondary: 37A50, 92C20 

\end{MSCcodes}

\section{Introduction}
A wide variety of physical scenarios involve nonlinear systems responding to external signals. Two mathematically well-studied classes of external signals include periodic signals and stochastic signals, among other types of stationary and potentially noisy signals. A particularly well-characterized scenario is the behavior of intrinsically oscillating systems responding to a periodic signal. In that domain, a variety of complex behaviors have been reported, including phase-locking patterns and chaos (e.g., \cite{glass1994}). 
Notably, excitable systems, although they do not produce oscillatory patterns on their own, can also exhibit similar phase-locking under supra-threshold periodic forcing \cite{feingold1988}. Stochastic forcing, classically seen as noise and often resulting in generating irregular behaviors, can in fact also, in excitable systems in particular, generate almost-periodic patterns through coherence resonance and self-induced stochastic resonance~\cite{pikovsky1997,lindner1999,muratov2005,deville2005}.
In physical or biological settings, it would be quite natural to expect that both periodic forcing and stochastic fluctuations are present.  For example, ecosystems and populations experience periodic seasonal forcing along with various forms of stochasticity, while physiological systems are subject to cyclical circadian inputs and other rhythmic signals in addition to diverse fluctuations.  The question we investigate here is how the combination of subthreshold periodic forcing together with stochasticity impacts dynamics, specifically in excitable systems. This question is less well studied, and appreciating heuristically the phenomena arising in these systems as a function of the forcing properties is complex.  
On the one hand, since both forms of forcing can induce regular responses, it may seem natural that they can work together to promote regularity.  On the other hand, they induce regularity through different mechanisms, with no guarantee of synergy. Moreover, it is notable that ecological fluctuations, from forest fires to hurricanes to disease outbreaks, although they generally arise within certain specific seasons, do not typically recur on a regular schedule or every year. 

A second, specific motivation for studying this scenario comes from auditory neuroscience. It has been well established that medial superior olive (MSO) neurons in the mammalian brain stem are responsible for sound localization. They receive binaural input, and reportedly achieve sound localization by producing spikes, or action potentials (large amplitude deviations of membrane potential away from their resting level), at temporally precise phases within the periodic sensory signals that drive them, despite jitter in the arrival times of these signals \cite{carr2010,schnupp2009}. Yet past work has shown numerically that depending on the variant of a computational MSO model that is used, this response precision may or may not occur \cite{meng2012,huguet2017}.
Thus, we set out to better understand the factors, from a dynamical systems perspective, that determine this outcome. More generally, we consider in this work the induction of large-scale events in excitable systems in response to the superposition of a subthreshold input, typically periodic, and phase-independent noise, and explore in particular the temporal precision of large-amplitude responses such as spikes, relative to the phase of the signal applied.

A substantial body of past work on excitable systems subject to noise~\cite{lindner2004effects} and constant input has provided a detailed understanding of various types of stochastic resonances and most likely trajectories that are associated with spikes~\cite{newby2014spontaneous,bressloff2014path,keener2011perturbation}. Spiking phenomena in response to inputs were also distinguished, in the context of Hodgkin's classification of neural excitability~\cite{prescott2008,hodgkin1948}, into three classes according to the responses induced by steps of applied current. Type 1 and Type 2 responses both involve a minimal current above which tonic firing will occur, but differ in their input-frequency response. In models, these properties are well described through the presence of specific types of bifurcation~\cite{rinzel1998,prescott2008}. Alternatively, Type 3 neurons maintain a stable equilibrium at any input level, and respond to steps of currents with transient, phasic spikes. In contrast to the other two response types, where equilibria disappear, the behavior of a Type 3 system is not well accounted for by bifurcation theory. Instead, spiking in these systems  has been characterized based on curves, also called quasi-threshold separatrices (QTS)~\cite{fitzhugh1961,prescott2008}, that separate regions of the phase plane associated with spiking or rest. Trajectory crossings of the QTS are often used as a proxy for spike generation, and understanding of the location of the QTS relative to orbits has been crucial to the analysis of trajectories and spiking in stochastic, excitable systems~\cite{muratov2005}. 

The understanding of neuronal responses to specific, time-dependent input profiles is much less understood. This scenario is especially central for the analysis of dynamics of Type 3 neurons, for which levels of static input do not directly translate into a spike rate. For Type 3 neurons, spiking depends on how the dynamical properties of input signals combine with the timescales and geometry of the system's orbits to shape an effective spike threshold. In particular, Type 3 neurons can selectively respond to specific slopes in the input and show post-inhibitory facilitation~\cite{rubin2021}.  The relationship of responses and inputs becomes even more complicated when the deterministic input is subthreshold and must be combined with stochastic effects to elicit a spike.  In~\cite{huguet2017}, it is reported that a model of Type 3 neurons of the MSO exhibited a tighter response precision to subthreshold periodic input together with noise than an associated Type 2 model of MSO~\cite{meng2012,huguet2017}. Here, we investigate the dynamical basis of these observations. Because these are non-autonomous systems, the QTS and quasi-static approximations of the QTS unfortunately fail to capture the relevant dynamics of the system, leading us to generalize the notion of threshold to dynamically varying inputs, beyond what was done previously for sequences of input steps~\cite{rubin2021}. 

The remainder of this paper is organized as follows.  In Section  \ref{sec:phaselock}, we introduce the two-dimensional auditory neuron model that will serve as a reference throughout this study and provide numerical illustrations of the variability in its response precision. Section \ref{sec:FES}  introduces a new construct, the dynamic threshold curve (DTC), which captures the phase-dependence of spiking to subthreshold deterministic input in the presence of noise. We also propose and discuss quantitative measures of phase locking when the input in this setting is periodic, which will allow us to quantify and compare {\em response precision} in various models. 
Once the DTC has been defined, it leads to two key questions:  (1) Why is the DTC useful for predicting and explaining the range of input phases over which neuronal spikes occur and the degree of phase locking that these responses exhibit?  (2) What properties of underlying dynamics and input determine the shape of the DTC?  
In Section \ref{sec:stochastic}, we address the first of these questions.
Specifically, we introduce a noise process that represents the distribution of trajectory positions in the presence of subthreshold periodic inputs and fluctuations, and we relate this construct to the DTC to generate an accurate approximation of the distribution of spike times relative to the periodic input signal.  We turn to the second question in Section \ref{sec:shape}, where we study how the features of the DTC are determined in a linear setting.   
In the linear case, we can precisely control the eigenvalues of the system's coefficient matrix as well as the deterministic trajectory's orientation to explore how these factors affect the dynamic threshold curve and response precision and we can also systematically study the effects of noise and threshold position on these features. Nonetheless, our simulations reveal a surprising lack of a straightforward relationship between local dynamic properties and the DTC.   Finally, in Section \ref{sec:homotopy}, we return to the nonlinear, two-dimensional neural model and introduce a homotopy of models that smoothly interpolates between the Type 2 and Type 3 dynamic regimes.  We show how varying the homotopy parameter affects the DTC and spiking probability as a function of periodic input phase and that the DTC indeed captures the previously observed difference in phase-locking between the two extreme cases.  We conclude the paper in Section \ref{sec:discuss} with a discussion that summarizes our results and presents some related open questions.

\section{Phase-locking in an auditory neuron model}
\label{sec:phaselock}
In 2012, Meng and collaborators~\cite{meng2012} introduced a planar model, dubbed the $V-U$ model, that represents a reduction of an 8-dimensional biophysical model for MSO neuron activity in the auditory brainstem \cite{rothman2003}. The $V-U$ model is  a conductance-based system of nonlinear ordinary differential equations, with a voltage variable $V$, which evolves in response to a sodium current with an inactivation variable $U$, low- and high-threshold potassium currents, a leak current, and an additional  hyperpolarization-activated cation current, coupled to the scalar, relatively slow $U$ equation:
\begin{equation}
    \label{eq:VU}
    \begin{array}{rcl}
    C_m dV & = & \displaystyle{\Big[I(t)-g_{Na}\,m_{\infty}^3(V)\,\frac{aU}{b}\,(V-E_{Na})-g_{KLT}(U)(V-E_K)-} \\
    & &\\
    & & \displaystyle{g_{KHT}(0.85n_0^2+0.15p_0)(V-E_K)-g_{L}(V-E_L)}\\
    & &\\
    & & \displaystyle{-g_h r_0 (V-E_h)\Big]\,dt + \sigma dW_t} \vspace{0.1in} \\
    & & \\
    dU & = & \displaystyle{\frac{U_{\infty}(V)-U}{\tau_U(V)}\,dt}
    \end{array}
\end{equation}
where $W_t$ is a Brownian motion. The parameters and functions appearing in system (\ref{eq:VU}) are described in detail in Appendix~\ref{sec:VUFunctions}, with default parameter values given in Table \ref{tab:VUparam}.  In these equations, the term  $I(t)$ is a time-dependent applied current, and $\sigma$ is the amplitude of the noise fluctuations. 

In line with~\cite{meng2012}, we investigate the \emph{response precision} of the neuron subject to a subthreshold input by considering the distribution of spike times for $\sigma>0$. A neuron is considered to have a high response precision if spikes, when they happen, arise within a narrow window of time with high probability (a phenomenon referred to as \emph{phase-locking} in~\cite{meng2012}); we will introduce specific tools to quantify response precision in Section \ref{sec:FES}.  
To describe the mathematical structures controlling the response precision of spiking, we consider the case where, in the absence of applied current and noise ($I(t)=\sigma=0$), system (\ref{eq:VU}) has a stable critical point at a relatively low $V$ value, corresponding to an inactive state (as is the case for the parameter values in Table \ref{tab:VUparam}).
Along the same lines as in~\cite{meng2012}, we will consider $I(t)$ to be a subthreshold, time-dependent signal, such that no spike is fired when $\sigma=0$.

\begin{figure}[h]
\centerline{\includegraphics[width=\textwidth]{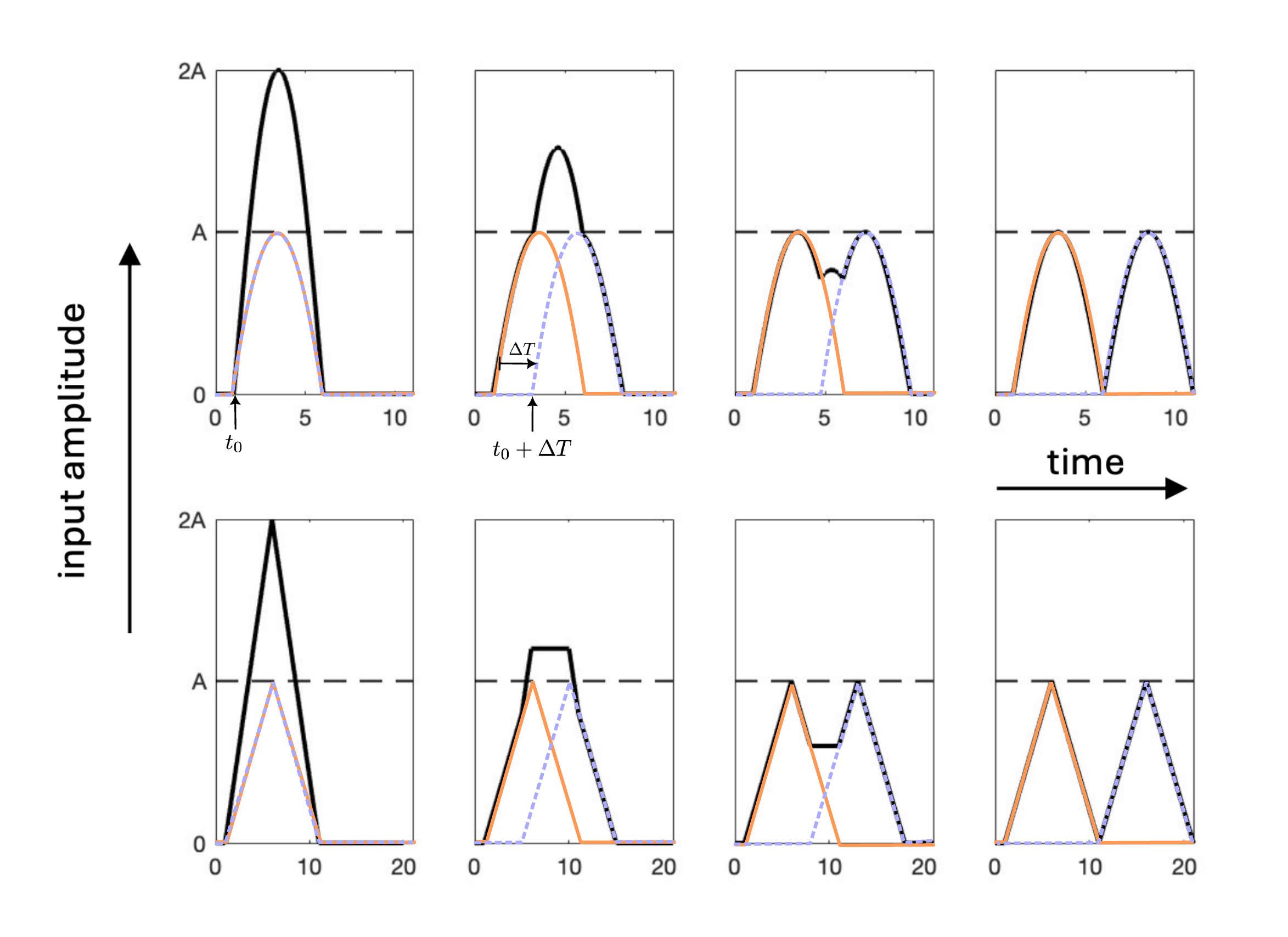}}
\caption{Examples of one period of the superposition of two time-shifted (solid orange and dashed blue curves) rectified sinusoidal inputs (top) and tent inputs (bottom), with $t_0=1$ and various $\Delta T$ (top: $\Delta T=0, 2.25, 3.75, 5$; bottom: $\Delta T=0, 4, 7, 10$). }
\label{fig:inputfig}
\end{figure}

In auditory signaling,  the neurons represented by system (\ref{eq:VU}) receive inputs from both ears with a timing difference that depends on the location of the sound source.  In addition to its biological relevance, the presence of this timing difference provides an interesting scenario for analysis, and hence we include it here.  
Specifically, we take $I(t)$ to be either a superposition of rectified sine waves or of linear tents (Figure \ref{fig:inputfig}), i.e. $I(t)=f(t-t_0)+f(t-t_0-\Delta T)$ with $t_0\geq 0$, where $f$ can take two forms: $f(t)=A\sin^+(2\pi\omega t)$ with $\sin^+(x)=\max(\sin(x),0)$ or $f(t)=\textrm{Tent}(t \ \mathrm{mod}\ \frac{2A}{\beta})$ with:

\begin{equation}\label{eq:tent}
\textrm{Tent}(t):=\begin{cases} 
\beta t & \text{for} \ 0<t<\frac{A}{2\beta},\\
\beta(\frac{A}{\beta}-t) & \text{for} \ \frac{A}{2\beta}\leq t \leq \frac{A}{\beta},\\
0 & \text{otherwise}.
\end{cases}
\end{equation}

\begin{figure}[!h]
\centering
\begin{overpic}[width=0.45\textwidth]{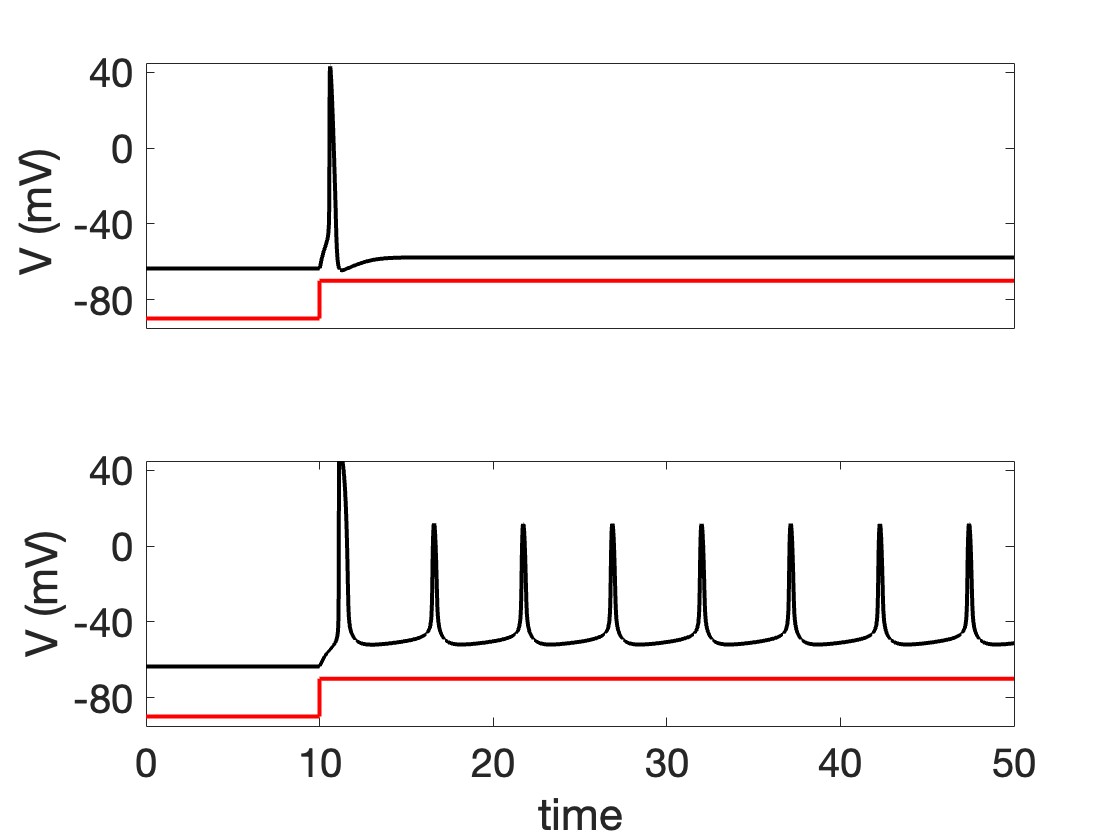}
    \put(-4,72){\large \textbf{(A)}}
\end{overpic}
\begin{overpic}[width=0.45\textwidth]{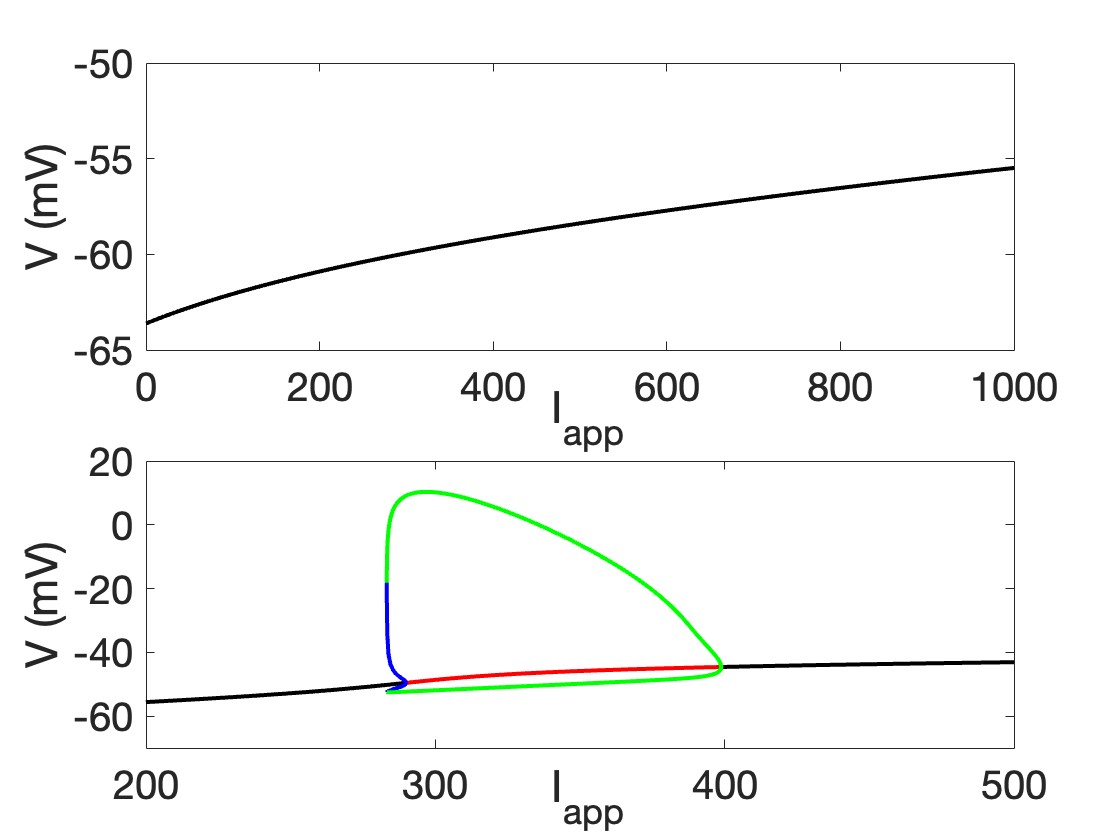}
    \put(-4,72){\large \textbf{(B)}}
\end{overpic}

\begin{overpic}[width=0.45\textwidth]{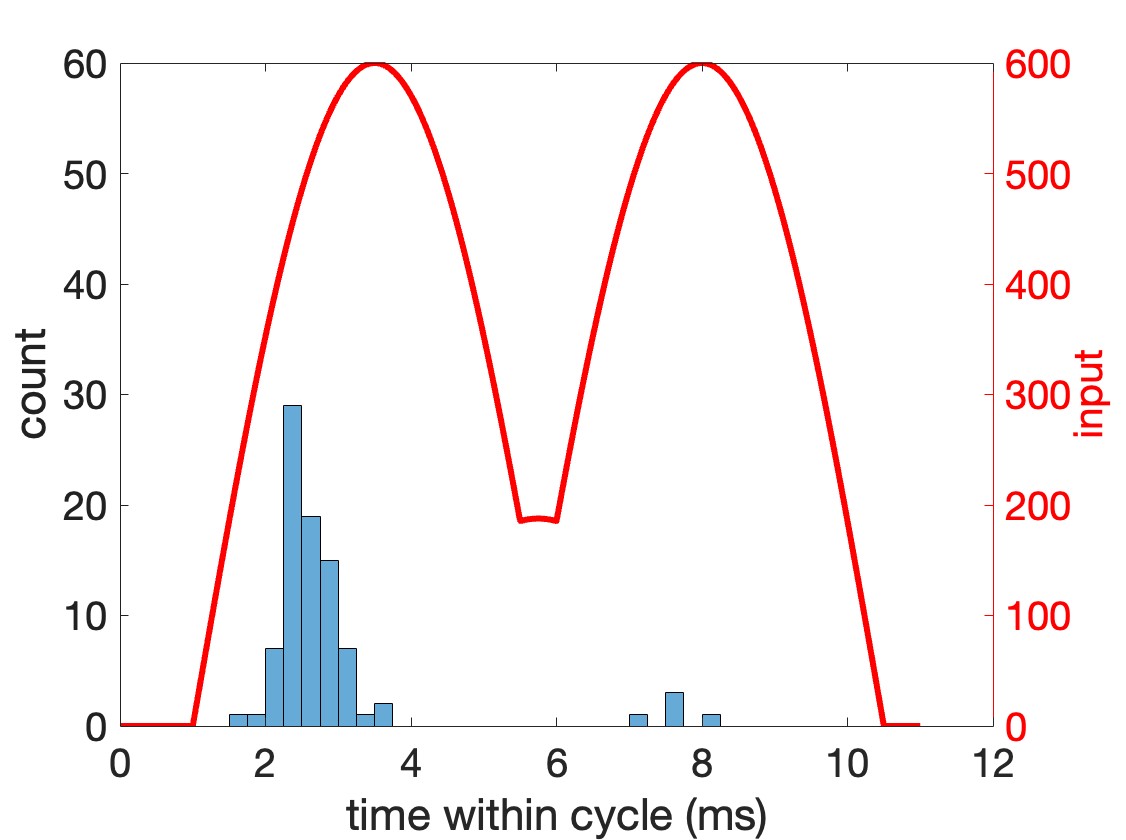}
    \put(-4,72){\large \textbf{(C)}}
\end{overpic}
\begin{overpic}[width=0.45\textwidth]{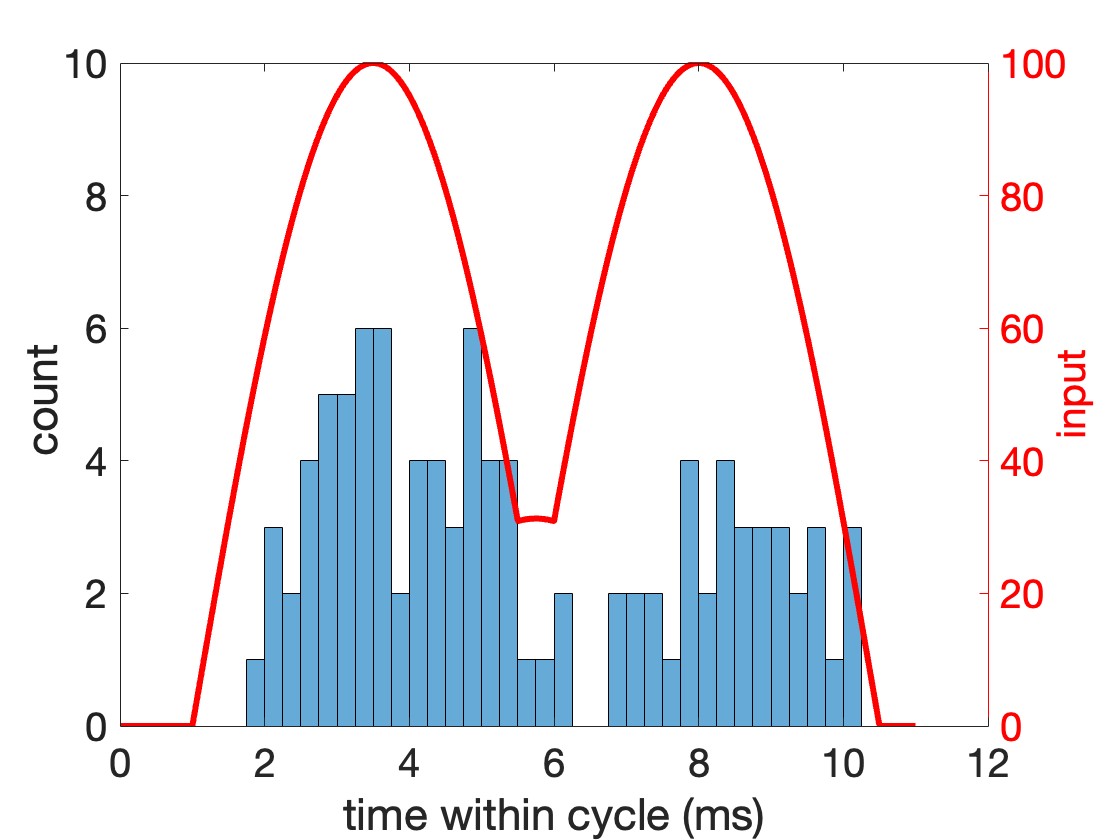}
    \put(-4,72){\large \textbf{(D)}}
\end{overpic}
\caption{Comparison of Type 3 and Type 2 responses.  (A)  Phasic or Type 3 (upper) and tonic or Type 2 (lower) voltage responses (black) to supra-threshold steps of input current (red) for the $V-U$ model (\ref{eq:VU}). (B)   Bifurcation diagrams for phasic (upper) and tonic (lower) cases with respect to applied current, $I(t)=I_{app}$. Stable (unstable) equilibria are shown in black (red), while maximal and minimal voltages of stable (unstable) periodic orbits are green (blue). (C)  Rectified sinusoidal input applied for multiple periods ($\Delta t=4.5$, period=11, resulting in 87 spikes) yields spikes within specific phases in the Type 3 case. 
(D) Rectified sinusoidal input applied for the same time duration ($\Delta t=4.5$, period=11, resulting in 98 spikes) yields spikes across a broad range of phases in the Type 2 case. 
Both regimes featured noise amplitude $\sigma=7.0$.  Parameters were adjusted to accommodate the increased sensitivity of the tonic model: reduced input amplitude, $g_{KLT}^{ton}=9.17$ and $g_{Na}= 1500$. } 
\label{fig:type3type2}
\end{figure}

System (\ref{eq:VU}) includes a commonly-used reduction in which the gating variable $U$ for the sodium current is also used to modulate the conductance of a potassium current, 
through a $U$-dependent conductance $g_{KLT}(U)$. This function controls the excitability type of the system, as described in~\cite{meng2012}:
\begin{itemize}
    \item In the {\em phasic} version of the model (i.e., a regime with Type 3 excitability), the low-threshold potassium conductance is a non-constant function $ g_{KLT}(U) \equiv g_{KLT}^{pha}(U):= \bar{g}_{KLT}a^4(1-U)^4z_0$, such that $-g_{KLT}(U)(V-E_K)$ is negative over the relevant $V$ range and grows in magnitude as $U$ decays, providing a negative feedback mechanism. With this feature, the model has a stable critical point at an inactive level of $V$ for any fixed $I(t)$ and deviates only transiently from there if $I(t)$ is stepped to a higher level, in what is called a {\em phasic} response (Figure \ref{fig:type3type2}, top).
    \item In the {\em tonic} model variant (i.e., with Type 2 excitability), this potassium conductance is assumed to be much faster and is approximated by its equilibrium value $g_{KLT}(U) \equiv 
    g_{KLT}^{ton} := \bar{g}_{KLT}a^4(1-U_0)^4z_0$, a positive constant, where $U_0$ is the $U$-coordinate of the critical point of system (\ref{eq:VU}) with $I=0$. 
In this case, with suitable parameter tuning (e.g., we increase $g_{Na}$ by $50\%$ relative to its baseline value), a sufficiently large step in $I(t)$ can produce a sustained, periodic train of large-amplitude excursions in $V(t)$, known as {\em tonic} spiking (Figure \ref{fig:type3type2}, middle). 
\end{itemize}



An intriguing observation made in~\cite{meng2012} is that these two models show distinct profiles of the distribution of spiking times when $I(t)$ is subthreshold and $\sigma>0$.  Representative examples of this phenomenon, with rectified sinusoidal $I(t)$, are shown in Figure \ref{fig:type3type2} (bottom row), with the phasic or Type 3 case on the left yielding relatively tight phase-locking and the tonic or Type 2 case on the right producing spikes across a broad distribution of phases.  In the next section, we introduce a new concept, the {\em dynamic threshold curve}, which is of general utility for characterizing the relationship between input phase and spike probability.

\section{The dynamic threshold curve (DTC) for forced excitable systems} 
\label{sec:FES}
\subsection{Thresholds in autonomous excitable systems}
Excitable systems are classically defined as dynamical systems with a single, globally stable equilibrium and with two modes of responses to perturbations from rest: small perturbations lead to a bounded return to the equilibrium state, typically monotonic in terms of distance to the equilibrium, while large enough perturbations induce a large excursion away from the equilibrium before a return to the vicinity of the equilibrium point occurs. For homogeneous excitable systems, a threshold can be defined mathematically using the notion of asymptotic rivers \cite{letson2018}, an approach that we used successfully in a study of other properties of Type 3 neurons \cite{rubin2021}. These \emph{thresholds} are typically manifolds of codimension one in the phase plane that separate initial conditions corresponding to the two modes of return to equilibrium. Asymptotic rivers are defined as particular solutions of the autonomous differential equation satisfying a few specific properties, and remain difficult to generalize to non-autonomous systems. 

\begin{figure}[h]
\centerline{\includegraphics[width=\textwidth]{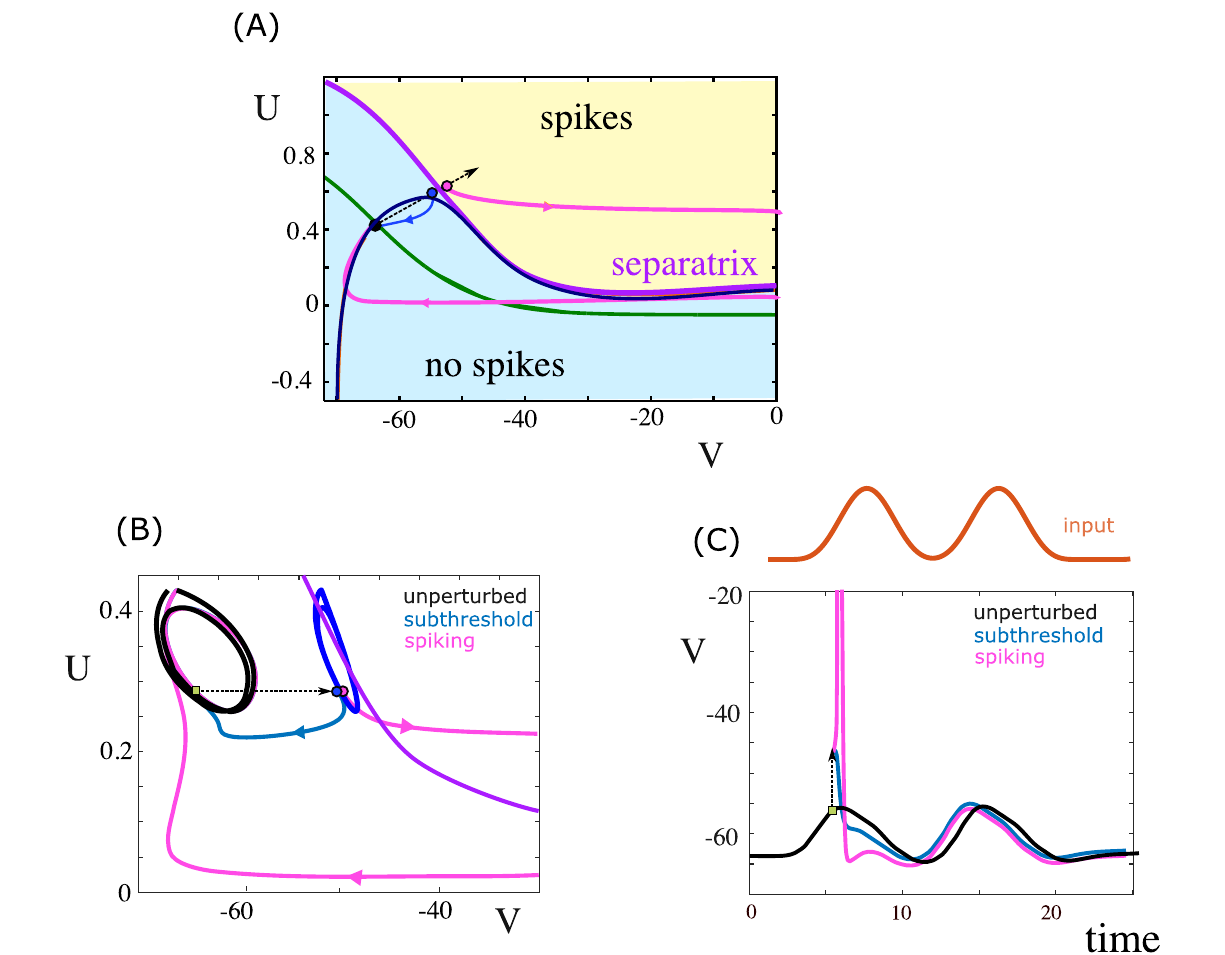}}
\caption{Thresholds in the $V-U$ model. (A) Static threshold (purple) separating a spiking region (yellow, with typical trajectory in pink, cut for legibility) from a non-spiking region (cyan region, with typical trajectory in blue). Black: $V$-nullcline, Green: $U$-nullcline, black circle: globally stable equilibrium point. Black arrow: direction of perturbation. (B,C) Dynamical threshold in response to a superposition of rectified sine waves (plotted in C, top). (B) Phase plane solution and (C) voltage time course, with the unperturbed but forced trajectory (black) plotted along with trajectories resulting from subthreshold (light blue) and superthreshold (pink) perturbations along the $v$-direction. The DTC (blue in B) lies between those two trajectories. The static threshold is plotted as a purple curve in (B) for comparison.}
\label{fig:thresholds}
\end{figure}

Figure~\ref{fig:thresholds}(A) illustrates the threshold for the $V-U$ model (\ref{eq:VU}) with $I(t)=\sigma=0$.  
In this system, as is generally the case in neuron models, excitability is interpreted as a natural property of neurons that, when stimulated strongly enough, generate action potentials that correspond to large excursions of the membrane voltage. In Figure~\ref{fig:thresholds}(A), the threshold, in purple, separates regions of the phase space where trajectories return to equilibrium without firing a spike (cyan region; e.g., blue trajectory) from regions of the phase space associated with spikes (yellow; e.g., pink trajectory). 

Defining a threshold is much more complicated for inputs that vary continuously in time than for instantaneous, fixed-amplitude inputs. Asymptotic rivers, which allow for the calculation of a threshold for each input level within a relevant range and thereby define a family of thresholds parameterized by input level (e.g., \cite{rubin2021}), typically cannot account for responses to dynamical inputs, whose variations in the future may facilitate or hinder a future spike firing. For example, with a rapidly changing input, even if a trajectory is initially above the static threshold associated with the instantaneous level of input at a given time (thus, on track to generate a spike if the input was to be suddenly frozen at its instantaneous level), a rapid decay in the input may cause the effective threshold to overtake the trajectory and the spike event might not occur. To deal with this situation, we introduce and make heavy use of what we call the {\it dynamic threshold curve} (DTC), which we now define. 


\subsection{Dynamic threshold curve (DTC) and dynamic threshold function: definitions and properties}\label{sec:DTC}
Consider the non-autonomous dynamical system
\begin{equation}\label{eq:generalDS}
\dot{x}(t)=F(x(t),I(t))
\end{equation}
where $x\in\R^d$ for some $d\geq 1$; $F:\R^d\times\R\mapsto \R^d$ is a smooth function, ensuring existence and uniqueness of solutions; and $I(t)$ is a $P$-periodic function of time (independent of $x(t)$) with $P>0$.  Denote by $X(t;t_0,x_0)= (x_1(t;t_0,x_0),\cdots,x_d(t;t_0,x_0))$ the unique solution of~\eqref{eq:generalDS} with initial condition $x_0$ at time $t_0$. 

Thresholds in excitable systems are defined relative to the occurrence of discrete events (large excursions away from the fixed point), and here we seek to establish a mathematical criterion to identify whether these events will happen as time evolves. To fix notation, let us define such events in terms of positive-direction crossings of a Poincaré section given by the hyperplane $\{x\in \R^d, x_1=0\}$, and focus on the case where $I(t)$ is a subthreshold input signal over the time interval $[0,P]$; that is, for all $t\geq t_0$, $x_1(t;t_0,x_0)<0$.
In line with our motivation in this work, let us consider the case where (\ref{eq:generalDS}) is subject to perturbations (typically, noise) that are one-dimensional and arise along a prescribed vector $p\in\R^d$.

\begin{Def}
\label{def:DTC}
The \emph{dynamic threshold function} $\kappa:[0,P]\mapsto \bar{\R}=[0,\infty]$ is the mapping that associates to $t \in [0,P]$ the minimal size of an instantaneous kick, applied to the deterministic trajectory at time $t$ along direction $p$, that will induce an eventual threshold crossing: 
\begin{equation}
    \label{eq:kappa}
\kappa(t)=\inf\{k>0 \; : \; \sup_{s\geq t} x_1(s; t, X(t;0,x_0) + k\, p) \geq 0\}.
\end{equation}
The \emph{dynamic threshold curve (DTC)} is given by the set of locations  in the phase plane attained by  application of each thus-computed minimal perturbation to the deterministic trajectory, which takes the form $\{ \Phi(t) : t \in [0,P] \}$  where for each such $t$, 
\begin{equation}
    \label{eq:DTCdef}
   \Phi(t) := X(t;0,x_0)+\kappa(t)\, p.
\end{equation}
\end{Def}
In particular, when $p$ is not parallel to the Poincaré section -- here, when $p\cdot e_1> 0$ where $e_1$ is the first vector of the canonical basis, the orthogonal vector to the Poincaré section considered -- the dynamical threshold function is well defined and finite for all time.\footnote{This notion lends itself to multi-dimensional generalizations. In particular, when considering multi-dimensional perturbations along dimensions $\{i_1,\cdots, i_m\}\subset\{1,\cdots,d\}$ at time $t$, a dynamical threshold could be defined as 
\[\phi(t)=\inf\{a>0 : \exists\; \tilde{X} \in B_m(X(t;0,x_0),a) \text{ and } \exists\; s\geq t \text{ s.t. } \tilde{x}_1(s)\geq 0 \}\]
where $\tilde{x}_1$ is the first component of the solution $\tilde{X}(s)$ of the differential equation starting at $\tilde{X}$ at time $t$ and $B_m(X(t;0,x_0),a)$ is the ball of radius $a$ centered at $X(t;0,x_0)$ with radius $a$ along directions $(i_1,\cdots, i_m)$; that is, it is the set of points $y\in\R^d$ such that $\sum_{k=i_1}^{i_m} (y_k-x_k(t;0,x_0))^2\leq a^2$ and $y_k=x_k(t;0,x_0)$ for $k\notin \{i_1,\cdots, i_m\}$. }
We emphasize the simple but crucial point that  $\kappa(t)$ and $\Phi(t)$ are not based on perturbations directly across the threshold but rather on minimal perturbations that yield eventual threshold crossing.

Definition \ref{def:DTC} has the advantage of being constructive and readily provides an algorithm for computing the DTC with arbitrary precision. For ease of presentation, we describe this algorithm for a two-dimensional dynamical system 
\[\begin{cases}
    v'=F_1(v,u)+I_u(t)\\
    u'=F_2(v,u)+I_v(t)
\end{cases}
\]
with $I(t)=(I_u(t),I_v(t))$ a $P$-periodic input and an initial condition at $t_0=0$ denoted by $x_0=(v_0,u_0)$. Let $v=v_{th}$ be the Poincaré section considered, and $p=(p_1,p_2)$ an arbitrary perturbation direction with $p_1>0$. An algorithm for computing the dynamic threshold function $\kappa(t)$ and associated DTC consists of the following steps, as illustrated in Figure~\ref{fig:thresholds}(B):
\begin{itemize}
    \item generate the deterministic trajectory of the model over one period of the input, $X(t; 0,x_0):= \{ \gamma(t)=(v(t),u(t)) : t \in [0,P) \}$ (black in Figure~\ref{fig:thresholds}(B));
    \item at each time $t^*\in [0,P)$ where the DTC is to be computed, we estimate $\kappa(t^*)$, up to a prescribed precision $\epsilon$, as the minimal value for $k$ such that $X(t; t^*,\gamma(t^*)+k\,p)$  crosses the threshold before a prescribed cut-off\footnote{In theory, there may be a trajectory that is asymptotically tangent to the threshold. Thus, for computational practicality, we impose an upper bound on how long we evolve perturbed trajectories before concluding that a threshold crossing has not occurred.} time $T>P$, and define the DTC via addition as in equation (\ref{eq:DTCdef}) (Figure~\ref{fig:thresholds} shows such perturbed trajectories for two different choices of $k$, with the blue below threshold and the pink above threshold).
\end{itemize}
In practice, to compute $\kappa(t)$ and the DTC with precision $\varepsilon$, various methods are possible, including brute-force algorithms that compute solutions of the system for each $k=n\varepsilon$ with $n\in \mathbb{N}$ increasing until a spike is recorded, or more efficient methods such as the dichotomic method described in Appendix~\ref{sec:numerics}, which was used for this paper.

The DTC provides a quantification of the sensitivity of the system to perturbations in a specific direction, taking into account the time-dependent input component of the system. 
We note that with these definitions, the dynamic threshold depends on the underlying vector field $F=(F_1,F_2)$, on the position of the trajectory at the time $t$ of each perturbation, and on the subsequent values of the input after that time, $\{I(s),s\geq t\}$. 
In Figure~\ref{fig:thresholds}(B), we observe that the DTC (blue curve) varies around the static threshold (purple curve); the shape of the DTC depends on the past and future of the input as well as the intrinsic nonlinear dynamics of the system. 

\begin{figure}[h!]
\centerline{\includegraphics[width=3in]{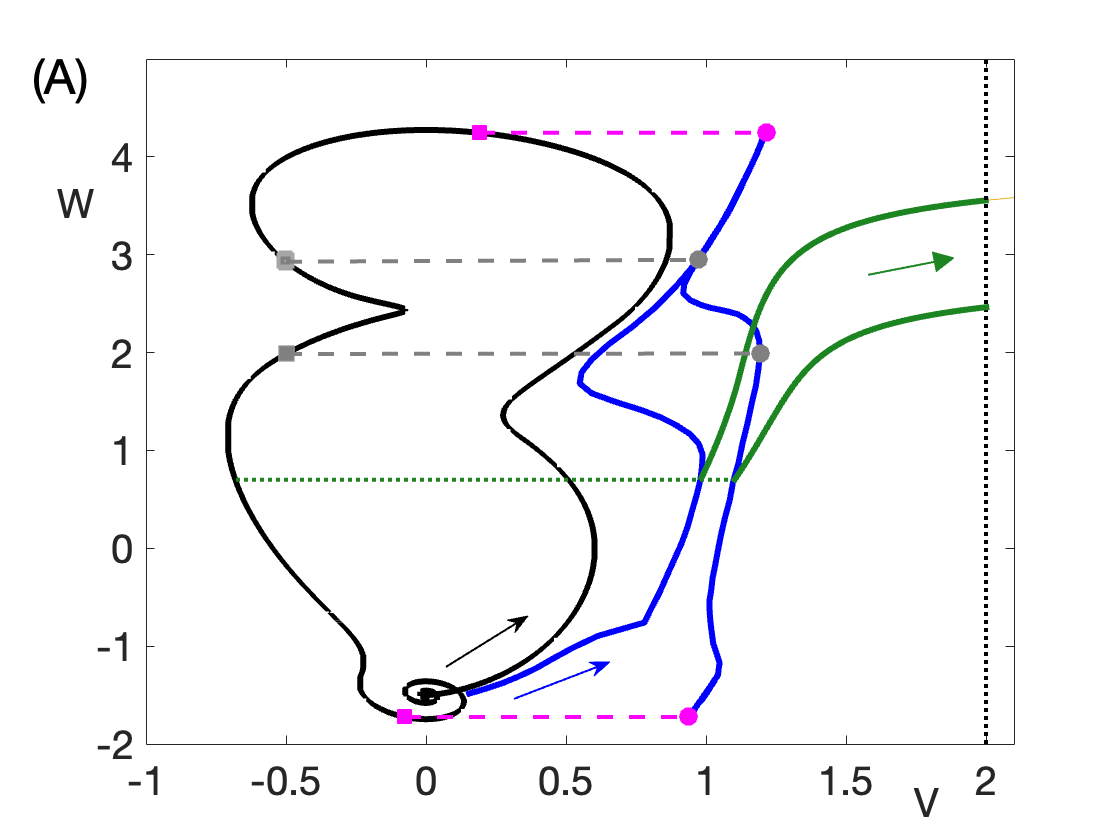}  \hspace{-0.25in}
\includegraphics[width=3in]{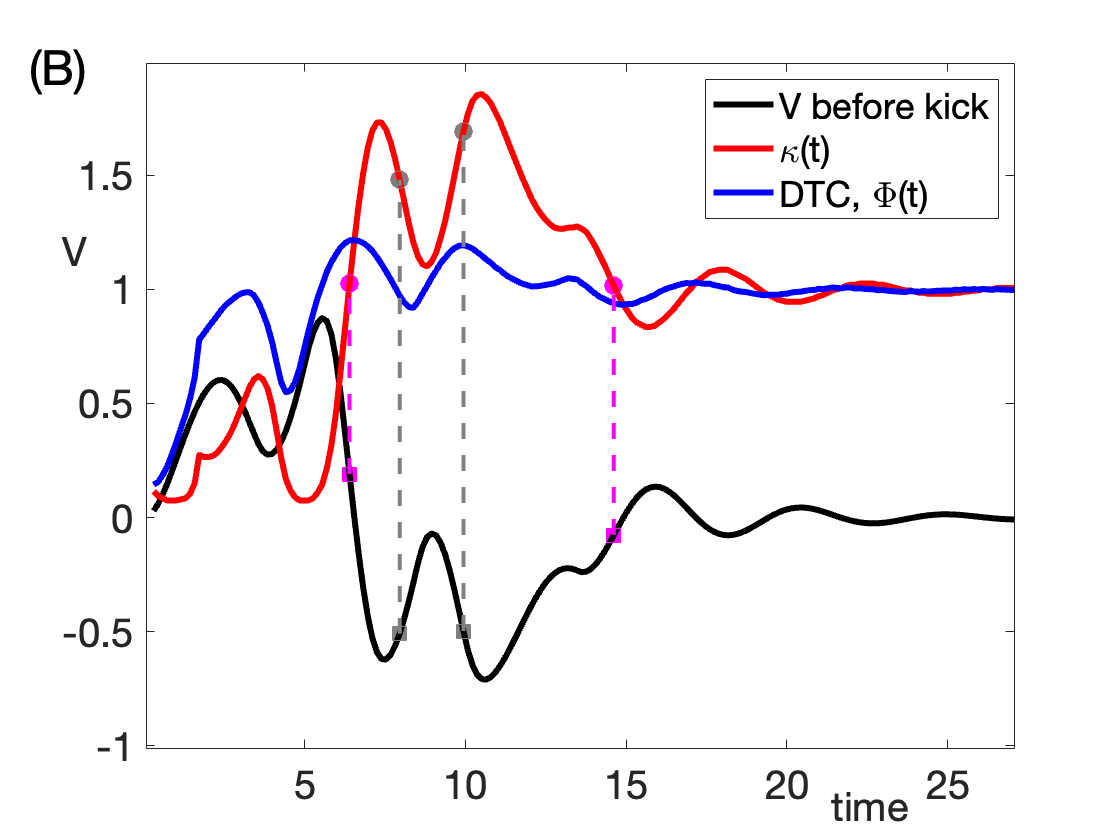}}
\caption{
Solutions of the quartic model~\eqref{eq:QIF} with tent input, together with the DTC (blue) and dynamic threshold function $\kappa(t)$ (red, on the right only). (A): Phase plane, with the trajectory (black; flow direction shown by arrow), position of the DTC (blue, with examples of perturbed trajectories exceeding the DTC depicted in green). (B): Voltage time course (black) with the DTC (blue) and $\kappa(t)$ (red) as functions of time. Note that (i) points with the same voltage may require different kick sizes to fire (gray lines), (ii) points with the same adaptation $W$ may require kicks to different voltages $V$ to spike (green line, left panel only), and (iii) the same kick size may be required for points with different $V$ coordinates (pink lines). Initial condition: $(V,W)=(0,-1.5)$ (stable fixed point), $\Delta t=3.5$.  Parameters:  $A=-0.25, B=2.0, C=0.0, D=1.0, \epsilon=1.0.$
}
 \label{fig:DTCdemo}
\end{figure}

Because the DTC depends on the system's intrinsic nonlinearities and future input levels, its shape is highly non-trivial. We illustrate these subtleties in Figure~\ref{fig:DTCdemo} in the case of the quartic integrate-and-fire model that we have analyzed in recent past works \cite{rubin2017a,rubin2017b,rubin2021}:
\begin{equation}
\label{eq:QIF}
    \begin{array}{rcl}
    V' & = & V^4 +2AV-DW+I_V(t), \vspace{0.1in} \\
    W' & = & \epsilon (BV-CW),
    \end{array}
\end{equation}
with parameters $A,B,C,D \geq 0$ and $\epsilon > 0$, with input $I_V(t)$ given by a pair of linear tent functions, as in equation (\ref{eq:tent}), and with a spiking threshold set at a fixed $V$ value, $\{ V = 2 \}$.
To generate the DTC, we start the model from its globally attracting critical point $(V_0,W_0)$ at $t=0$ (where $I_V(0)=0$) and apply the algorithm described above.  
Figure~\ref{fig:DTCdemo} represents the phase plane (left) and time trace of the voltage variable (right) for the system, with the solution of the forced quartic neuron (in black), the DTC (in blue), and the associated magnitude of the dynamic threshold function $\kappa(t)$ (red on the right figure). 
We see that the function $\kappa(t)$ varies significantly between very small values and values that are almost twice the voltage difference between rest and threshold, and reflects both previous and future variations in the input and the nonlinear neuron's dynamics. In coarse terms, periods of declining input tend to lead to lowered voltage values and a need for larger kick sizes, such that the peaks in $\kappa(t)$ align with troughs in $V(t)$.

We can also observe several interesting features of how the DTC relates to the deterministic trajectory $X(t)=(V(t),W(t))$.  Multiple, distinct DTC points can arise for the same value of $W$ along $X(t)$, as we illustrate by the dotted green horizontal line at $W=0.7$; this line intersects $X(t)$ twice, leading to two different points on the DTC (Figure~\ref{fig:DTCdemo}A).  This distinction can only happen because the system is not time-homogeneous, and it reflects the impact of the time course of the input. Indeed, the larger DTC value for this example corresponds to the point on $X(t)$ with smaller $V(t)$, since the trajectory must be kicked closer to threshold to spike from this low-input position.  Similarly, the squares on the two gray lines represent two points on $X(t)$ associated with the same voltage $V^*=V(t_1^*)=V(t_2^*)$ but distinct $W$ values; they map to distinct values of the dynamic threshold function $\kappa(t)$ and  to distinct $V$ values on the DTC (gray dashed lines to circles, Figure~\ref{fig:DTCdemo}A,B). Lastly, the two pink lines indicate points associated with identical distances from $X(t)$ to the DTC (i.e., identical values $\kappa(t)=1$, Figure~\ref{fig:DTCdemo}B), but they are clearly associated with quite distinct locations in the phase plane, with different values of both $V(t)$ and $W(t)$ (Figure~\ref{fig:DTCdemo}A), as well as to different $V$ coordinates on the DTC.

\subsection{Response precision} 

\begin{figure}
    \centering
    \includegraphics[width=\linewidth]{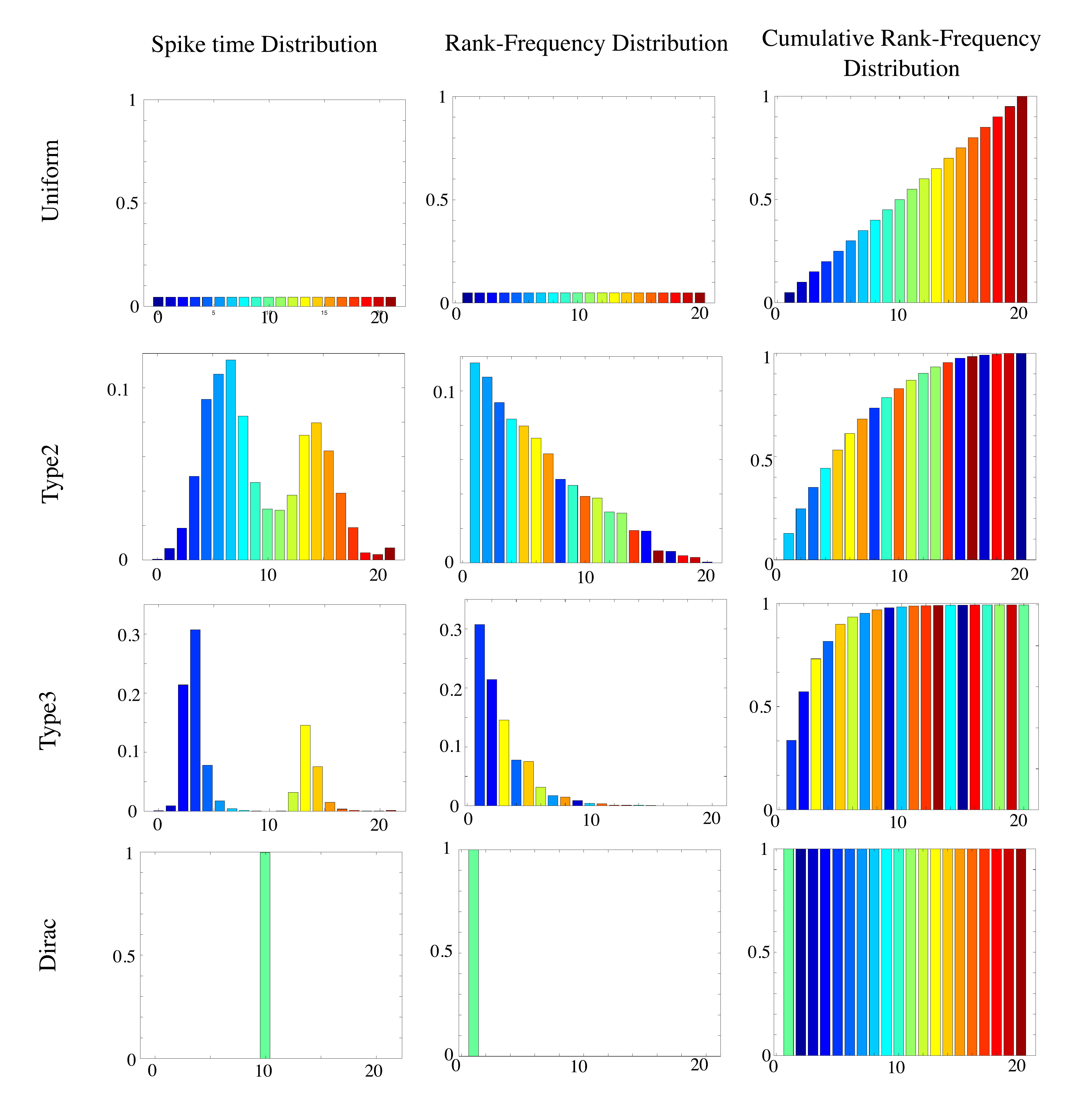}
    \caption{Response precision  for four distributions: from top to bottom, uniform, VU model in a type 2 regime, VU type 3 and a Dirac distribution. Left: distribution, middle: rank-frequency distribution and right: cumulative rank-frequency distribution. The color identifies the bin in the original distribution - and bars from the histogram conserved their original color after sorting them in the rank-frequency distribution, and when they are added to the cumulative distribution. Distributions were sorted according to increasing response precision, going from $0$ for the uniform distribution to 1 for the Dirac distribution. The type 2 neuron model was associated with a response precision of $0.50$ and, for type 3, $0.82$.} 
    \label{fig:RP_Principle}
\end{figure}

This study focuses on response precision, which corresponds to how synchronized, in time, are the responses of the forced excitable system when subject to noise. This notion is, in many aspects, similar to the classical phenomenon of phase locking in periodically forced systems. Classical phase locking to a signal corresponds to the phenomenon whereby a neuron fires (or a system oscillates) with a fixed phase shift relative to a periodic signal.  However, the spike distributions in stochastic systems and the type of multimodal distributions arising in auditory neuroscience models~\cite{meng2012}, as represented in Figure~\ref{fig:type3type2}, require a less strict or restrictive concept of phase locking.  Indeed, in stochastic systems, phase locking is only approximate; different models may present distinct spreads around a given phase shift, motivating the extension of the notion of phase locking to probability distributions of phase differences. Moreover, although the distributions observed may feature multiple peaks, they may nonetheless exhibit a high degree of phase locking around two or more phase shifts. To  quantify the degree of precision of responses in such situations, we developed a new measure of {\em response precision (RP)}. This measure relies on an analysis of the phase histogram that takes into account the dispersion of phases and allows for the possibility of  multiple peaks, while ignoring gaps between them. To realize this measure, we computed a large sample of the events of interest  across many realizations of the stochastic trajectories (or, when the input is periodic, multiple periods the signal, with time considered modulo the period $P$) and estimated the phases of these events. We also fixed a common discretization of the relevant time interval $[0,P)$ into $n$ equally spaced bins. From this collection of data, we estimated response precision as follows (see Figure~\ref{fig:RP_Principle}):
\begin{itemize}
\item we constructed the normalized (relative to the total number of events in the sample) histogram $(h_1,\cdots, h_n)$ of phases of the events;
\item we computed the rank-frequency distribution by sorting the bins according to their amplitude, starting from the bins associated with the largest number of events, to obtain a new discrete, decreasing histogram $(\tilde{h}_1,\cdots,\tilde{h}_n)$;
\item we computed the cumulative distribution associated with this distribution, $(c_1,\cdots,c_n)$ with $c_k=\sum_{i=1}^k \tilde{h}_i$; 
\item we estimated an unscaled response precision, uRP, as the area under the curve, $\sum_{k=1}^n c_k$;
\item finally, this measure was rescaled (see below) to give the RP:
\begin{equation}
    \label{eq:RP}
RP=\frac{2\,uRP-(n+1)}{n-1}.
\end{equation}
\end{itemize}

\begin{remark}
Alternatively, the RP can be seen as $n-\mathbb{E}[\tilde{X}]$ where $\tilde X$ is a discrete random variable on $\{1,\cdots,n\}$ with distribution $(\tilde{h}_1,\cdots,\tilde{h}_n)$. This identity stems from the classical relationship between the sum of the complementary cumulative distribution of a random variable and its expectation. 
\end{remark}
We note that the RP achieves a reasonable estimate of phase precision as understood in our context:
\begin{itemize}
    \item For unimodal distributions, the RP will reflect the spread of the phases around the mean, with the largest RP arising when all events happen in the same bin, and the smallest RP arising for a uniform distribution of phases. 
    \item For multi-modal distributions, the RP will ignore the location of the centers of the modes, but keep track of the dispersion around each of the peaks, as well as the relative probabilities in each of the peaks. In particular, the RP of a two-peak distribution with only a small fraction of events in the second peak will only be slightly larger than the RP of a similar distribution in the absence of the second peak. 
\end{itemize}
These observations lead to the scaling formula used in the definition of RP in (\ref{eq:RP}) from the uRP introduced above. First, it is clear that for a perfectly phase-locked system (at the resolution of the histogram considered), the rescaled histogram will have $h_k=1$ for some $k\in \{1,\cdots,n\}$ and $h_l=0$ otherwise, so $\tilde{h}_1=1$ and $\tilde{h}_l=0$ for $l>1$, and thus the cumulative distribution is uniformly equal to $1$, yielding a uRP equal to $uRP_{max}=n$. Alternatively, for the uniform distribution, both histogram $h$ and $\tilde{h}$ are uniform, $c_k=k/n$, and the uRP is equal to $uRP_{min}=\frac{1}{2} (n+1)$. The RP formula is thus given by $RP=\frac{uRP-uRP_{min}}{uRP_{max}-uRP_{min}}$, which yields (\ref{eq:RP}).

\subsection{Linking DTC and RP: organization of analysis}
With the DTC and RP defined, there are at least three key questions to consider, which we partially address in the remainder of this study: 
\begin{enumerate}
    \item From a dynamics perspective, what makes the
DTC a useful tool for predicting and explaining the range of input phases
over which neuronal spikes occur and the degree of phase locking
that these responses exhibit (see Section \ref{sec:stochastic})?
\item What properties of
underlying dynamics and applied input determine the shape of the DTC (see Section \ref{sec:shape})?
\item Which DTC properties determine when spiking occurs and what degree of phase-locking results (see Section \ref{sec:homotopy})? 
\end{enumerate}
For the final point, one would intuitively expect that spikes would be most likely to occur in the phases at which the deepest troughs of the dynamic threshold function used to compute the DTC arise, since these valleys correspond to the closest approach of the deterministic trajectory to the threshold and hence should mark positions from which the smallest noise-induced deviations are needed to achieve threshold crossing. 
Yet, other features could be equally or more important; for instance, given the same depth, two troughs could have different second derivatives and asymmetries that could affect the DTC, and it is not clear how the RP depends on such details of DTC shape nor how closely a peak in spiking probability will align with the actual minimum of the dynamic threshold function.  All of these factors may depend on parameters of the underlying deterministic model and input function, as well as the noise intensity.  

To study these relationships, we will compare features of the DTC (or dynamic threshold function) and RP under the systematic variation of certain input and model features.  
Specifically, we will record spike times during long simulations involving periodically repeating inputs, with noise present.
We note that one potentially confounding factor in this study could be the difference between what happens to trajectories after spikes versus how trajectories behave on cycles when they do not spike.  That is, trajectories after spikes could be at different locations in the phase plane during subsequent input phases than the locations where the non-spiking trajectories lie, leading to different subsequent spike phases.  To reduce this effect, we will consider systems (\ref{eq:generalDS}) in regimes in which there is a globally attracting equilibrium point $x_0$ for $I(t) \equiv 0$.  Although we will consider nonzero choices of $I(t)$ with period $P>0$, these will be scaled such that the resulting deterministic trajectory $x(t)$ does not spike on $[0,P]$. 
When noise is introduced, it will be scaled in such a way that spikes do occur on some cycles.  Finally, we will choose input patterns with $I(t)=0$ on the later part of the $[0,P]$ time interval, such that spiking and non-spiking trajectories have time to settle near the equilibrium point in between input cycles in both cases.

Finally, note that when noise is included, another form of locking or reliability could relate to the variability in terms of numbers of input cycles, or time intervals of length $P$, skipped between spikes; that is, firing spikes once every $n$ cycles for some fixed $n$ could be viewed as a more locked or reliable response than spike emission with a more diverse distribution of skipped cycle counts.  Under the assumption of return to the same starting point after each cycle, however, such that the probability of spiking on any cycle can be taken as a constant $\varphi$, the variance in cycle skipping can be computed as follows.  Let $X$ be a random variable representing the number of cycles skipped between spikes.  We have $\E[X^k] = \sum_{n=1}^{\infty} n^k \varphi (1-\varphi)^n$  and hence var($X$) = $\E[X^2]-\E[X]^2 = (1-\varphi)/\varphi^2$.  Thus, as long as we vary the noise intensity to keep the cycle-by-cycle spike probability roughly constant for inputs with different features, this variance will not change.  Hence, we will not consider this form of variability in the rest of this work.

\section{The dynamic threshold function $\kappa$ predicts stochastic spike timing distributions}
\label{sec:stochastic}


We return here to our original motivation of characterizing spike timing for a stochastic system
\begin{equation}\label{eq:generalSDE}
dy=F(y,I(t))\,dt+\Sigma \cdot dW_t
\end{equation}
where $y\in\R^d$ is the stochastic counterpart of the variable $x$ in equation~\eqref{eq:generalDS}. In all generality, $W_t$ is a $k$-Brownian motion and $\Sigma$ a $d\times k$ matrix, but for simplicity of the exposition, we assume here, in line with the assumptions on~\eqref{eq:VU}, that $k=1$ and denote $\Sigma=e\in\R^d$, so the noise perturbations are one-dimensional.  We will also take $p=e$, such that the input is applied to the same equation in (\ref{eq:generalSDE}) where the noise has a non-trivial contribution. Section~\ref{sec:DTC} proposed to estimate the minimal amplitude $\kappa(t)$ of a deterministic perturbation along vector $e$, applied at time $t$, that would lead the system to spike. It is thus natural to assume that spikes in the stochastic system (\ref{eq:generalSDE}) happen, in first approximation, when $y(t)$ crosses the DTC $\Phi(t) =  x(t)+\kappa(t)\,e$. We test this hypothesis in this section, further reducing the complexity of the system by focusing our attention on the crossings of $\kappa(t)$ by the \emph{fluctuation process} $\delta(t)=(y(t)-x(t))\cdot e$  that describes the deviation along the vector $e$ between the stochastic and the deterministic trajectory. 

We assume that spikes are faster than the subthreshold neural dynamics or, more precisely, we ignore the delay between spike triggering and the effective spike emission; we also neglect the possibility that noise could  prevent the neuron from spiking even if the dynamic threshold were crossed. In this case, spike times correspond to first passage times of the stochastic process $\delta(t)$ to the dynamic threshold function $\kappa(t)$:
\[\tau=\inf\Big\{t\geq 0, \delta(t)\geq \kappa(t)\Big\} \equiv
\inf\Big\{t\geq 0, x(t)+\delta(t)\,e \geq \Phi(t) \Big\}. \]


This quantity is a random variable since $\delta(t)$ is a stochastic process. The problem of identifying the spike time of a stochastic, forced system thus reduces, in this approximation, to finding the probability distribution of first passage times of the  process $\delta(t)$ to a nonlinear boundary. First passage time problems are classical problems in stochastic calculus, for which no systematic solution exists for general processes and nonlinear  
boundaries~\cite{karatzas2014brownian,redner2001guide}. 
Indeed, the fluctuation process $\delta(t)$ is a  solution of the equation: 
\[d\delta(t)=(F(y(t),I(t))-F(x(t),I(t)))\,dt+\Sigma \cdot dW_t.\]
This process is typically non-Gaussian, non-centered (due to the nonlinear, excitable dynamics), and non-stationary (owing to the time-dependent input),  properties that make the derivation of first passage times to $\kappa(t)$ out of reach of usual analytical methods\footnote{While the theory of first hitting times of one-dimensional stochastic processes has been the object of a broad literature, analytical results for the first hitting time of a general process to a time-dependent boundary remain scarce and fragmentary, and mostly applicable to simple diffusion processes that are either Gaussian or reducible to Gaussian processes. For such processes, analytical methods include those relying on martingale problems (e.g., reflection principle, ~\cite{karatzas2014brownian}), Girsanov theorem~\cite{touboul2008characterization}, or reductions to integral equations~\cite{buonocore,durbin-williams:92,durbin:85,durbin:92} or to partial differential equations with complex boundary conditions (Feynman-Kac techniques~\cite{bass,ito-mckean,stroock-varadhan}). Asymptotic approaches and moment estimates have been widely used to characterize properties of first hitting times~\cite{nobile-ricciardi-etal:85,nobile-ricciardi-etal:85b}, and a variety of numerical methods based on Monte-Carlo simulations or more sophisticated numerical probability approaches~\cite{taillefumier2011multiresolution} have also been proposed (see also review in~\cite[Chapt. 6,7]{touboul2008nonlinear}).}.

\begin{figure}
\includegraphics[width=\textwidth]{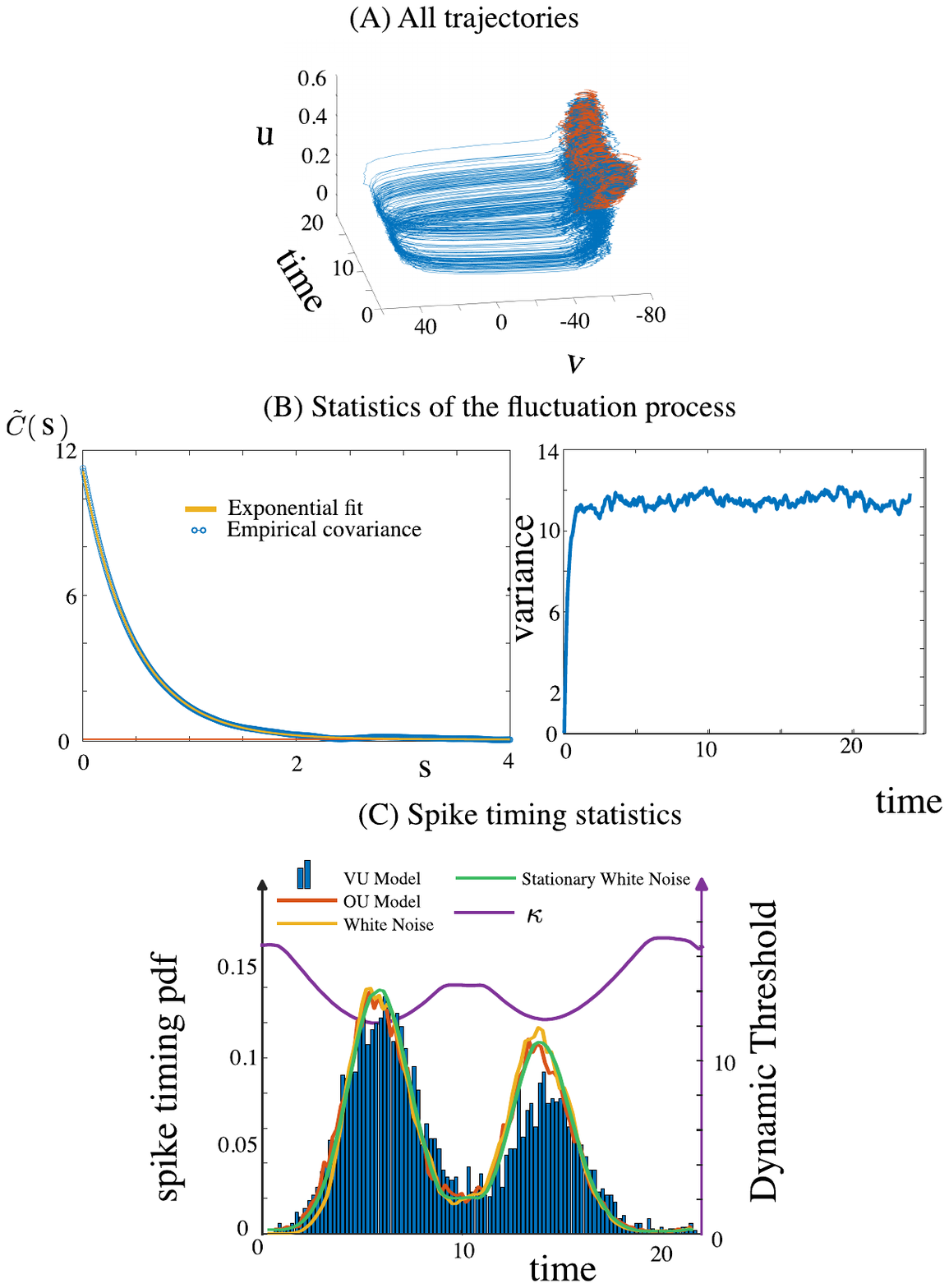}
\caption{Type 2 neurons with noise. (A) 100 sample trajectories, some spiking (blue) and some not (red). (B) Time-averaged empirical covariance, $\tilde{C}(s)$ (left, with exponential fit in yellow) and instantaneous variance (right)  
computed from the non-spiking trajectories.  (C) Spike time statistics for the $V-U$ model (blue histogram), together with the distributions of the first hitting times to the dynamic threshold function (purple) for the Ornstein-Uhlenbeck model (red), the white noise model (yellow), and the stationary white noise (green). All three models show a strong consistency.}
\label{Type2-Sto}
\end{figure}

\begin{figure}
\includegraphics[width=\textwidth]{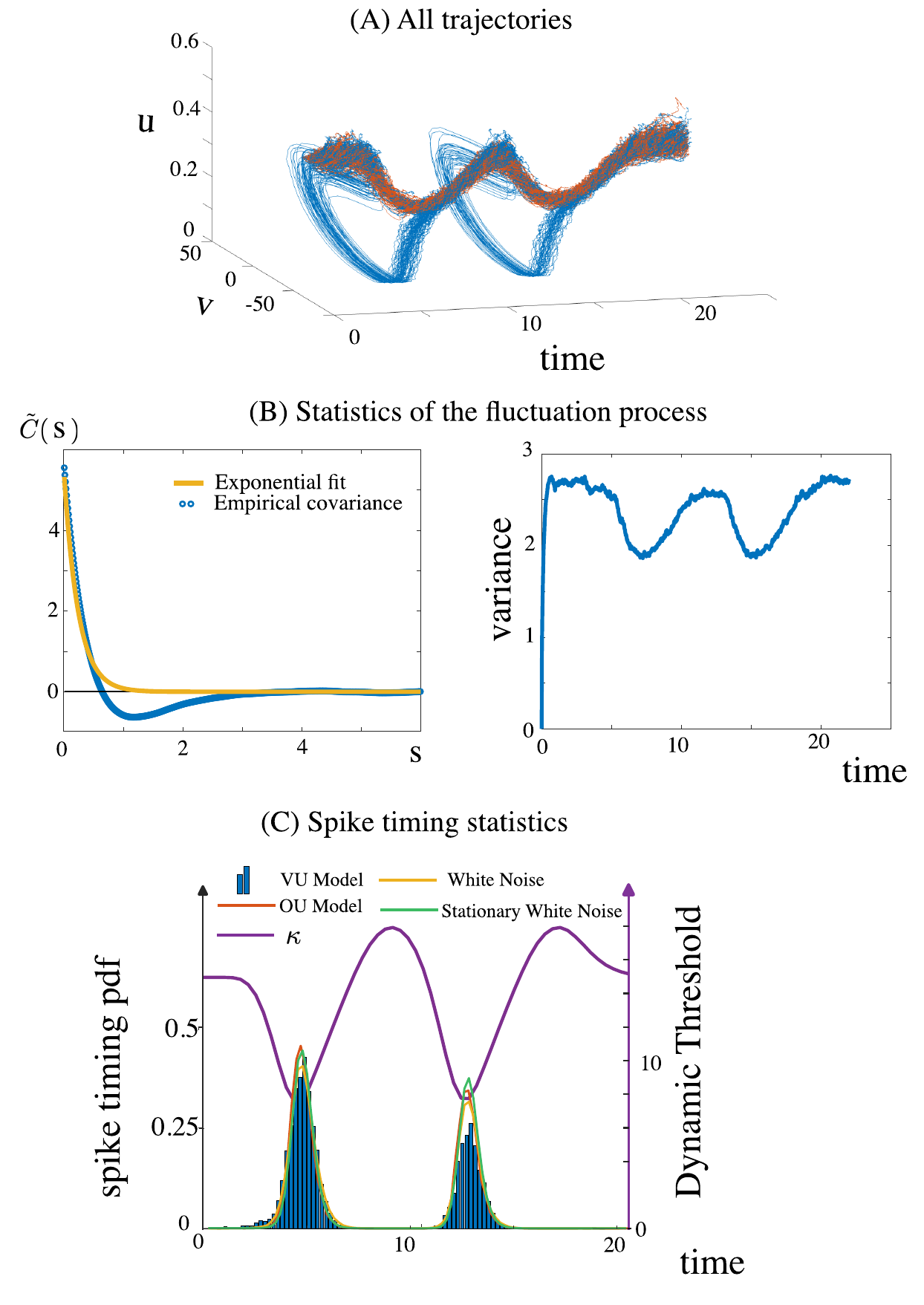}
\caption{Same as Fig.~\ref{Type2-Sto} for Type 3 neurons.}
\label{Type3-Sto}
\end{figure}

In order to investigate whether the dynamic threshold function $\kappa(t)$ indeed provides information about the probability distribution of spike times, we introduced two approximations of the fluctuation process $\delta(t)$ and used these to compute the distribution of the associated first hitting times to $\kappa(t)$.
We first estimated a stationary Ornstein-Uhlenbeck process
\begin{equation}\label{eq:OU}
dX_t=-\frac{1}\tau X_t+\sigma dW_t,
\end{equation}
where $\tau$ and $\sigma$ were fitted so the autocovariance of $X_t$ matches the statistics of the fluctuation process $\delta (t)$. The autocovariance of $\delta$ was computed using a Monte-Carlo simulation in which  a large number of realizations of the stochastic $V-U$ model were computed (see Figure~\ref{Type2-Sto}A and~\ref{Type3-Sto}A for Type 2 and Type 3 neurons, respectively). Trajectories were split into two categories: those that were associated with a spike (blue trajectories), which were used to compute the distribution of spike timings and the probability to spike, and those that did not spike, the statistics of which provided information about the statistics of the fluctuation process conditional on the trajectory remaining subthreshold (red trajectories). The latter trajectories were used as a sample of the fluctuation process and hence to compute statistics to fit our Gaussian models; specifically, the sample-averaged (or empirical) covariance of the fluctuation process,  $C(t,t+s):=Cov(\delta(t),\delta(t+s))$, was extracted. Since the process is not stationary, this quantity depends on time $t$. We thus also estimated the time-averaged empirical covariance, $\tilde{C}(s)=\frac{1}{T}\int_{t=0}^T C(t,t+s)\,dt$ (Figure~\ref{Type2-Sto}B and~\ref{Type3-Sto}B, left, blue circles). For our choice of parameters, we observed that Type 2 neurons displayed a time-averaged covariance $\tilde{C}(s)$ very well approximated by an exponential function, while the time-averaged covariance $\tilde{C}(s)$ computed for Type 3 neurons, although it too exhibited an exponential decay in amplitude, also featured damped oscillations. Recalling that the autocorrelation of the stationary Ornstein-Uhlenbeck process~\eqref{eq:OU} is given by $Cov(X_t,X_s)=\frac{\tau\sigma^2}{2}e^{-\vert t-s\vert/\tau}$, we estimated $\tau$ and $\sigma$ by fitting this function to the time-averaged empirical covariance function $\tilde{C}(s)$ (Figures~\ref{Type2-Sto}B and~\ref{Type3-Sto}B, left, yellow curve). We observed that equation (\ref{eq:OU}) with these parameters and with spike times defined by passage from 0 to $\kappa(t)$, despite being simple and linear, was sufficient to accurately predict the distribution of spike times of the $V-U$ model in both Type 2 and Type 3 regimes (red curve versus blue histogram in Figures~\ref{Type2-Sto}C and~\ref{Type3-Sto}C).

This clear agreement led us to seek a further simplified approximation to the fluctuation process. To this end, we generated realizations of a white noise stochastic process -- that is, a time series of a random variable with a Gaussian distribution of prescribed variance -- and computed the probability for this process to exceed the dynamic threshold function $\kappa(t)$. 
We used two models to fit the variance: one a non-stationary white noise model with a variance given by the estimation of the instantaneous variance $v_{\delta}(t)$ across stochastic $V-U$ model trajectories  conditional on not spiking (Figure~\ref{Type2-Sto}B and~\ref{Type3-Sto}B, right) and the other with constant variance, equal to the variance averaged across these trajectories over the whole simulation. In a time interval $[t,t+dt]$, the probability for the white noise process to exceed the threshold  $\kappa(t)$  is given by:
\[\rho(t)=\frac{1}{\sqrt{2\pi v(t)}}\int_{\kappa(t)}^{\infty} e^{-x^2/2v(t)} \, dx \]
where $v(t)$ is the variance of the process (either constant or $v_\delta(t)$). 
Given a fixed grid size $\varepsilon$, the probability that the first hitting time of the white noise process is $t=k\varepsilon$ is estimated as $\rho(k\varepsilon)\prod_{i=1}^{k-1}(1-\rho(i\varepsilon))$; see Figures~\ref{Type2-Sto}C and~\ref{Type3-Sto}C. Strikingly, we found that even the constant variance, stationary white noise process provides a very accurate prediction of the spike times, indicating that $\kappa(t)$ and hence the DTC itself indeed encapsulate critical information about the spike timings of the nonlinear model.  


The fact that the calculation of the spike time distribution in the stationary white noise model relies on a closed-form formula and does not require any precise estimates of the fluctuation process has significant computational and theoretical consequences. Computationally, this model enables efficient estimation of possible spike distributions for various noise levels, eliminating the need for costly Monte Carlo simulations. 
Theoretically, this finding shows that the DTC, which is computed from a deterministic trajectory, is a crucial quantity that contains most of the information relevant to spike timing, regardless of the properties, other than the mean variance, of the fluctuation process that allows trajectories of the stochastic system to cross the DTC. 

When we compare Type 2 versus Type 3 neurons across Figures~\ref{Type2-Sto}B and~\ref{Type3-Sto}B, we note two essential differences in the statistics of the fluctuation process: (i) the covariance function in our Type 2 model neuron is almost perfectly exponential and contrasts with the undershoot to negative values in our Type 3 model neuron, and (ii) the variance of the fluctuation process is essentially constant in our Type 2 model neuron, but shows substantive fluctuations with drops approximately coincident with the two troughs of the dynamic threshold function $\kappa(t)$ and the two peaks of the spike time distribution in the Type 3 case.  It appears that the reduced variance here, combined with the deeper wells of $\kappa(t)$, yields tighter phase-locking for the Type 3 regime. In the Type 3 case, with weaker fluctuations in the dynamics, the firing of spikes is limited to small neighborhoods of the troughs of $\kappa(t)$.  Despite the oscillations in the variance in this case (Figure \ref{Type3-Sto}B), the stationary white noise does almost as well as the time-varying white noise at fitting the histogram of spike times (Figure \ref{Type3-Sto}C) because the variance troughs align with the peaks of $\kappa(t)$, where spike probability is infinitesimal.  In contrast, in the Type 2 case, with a larger, relatively constant variance, the fluctuations can push the trajectory across the spike threshold over broader neighborhoods of the troughs of $\kappa(t)$, resulting in distinctions between phase-locking in the two cases that are even more significant than would already be predicted based on comparison of trough shapes alone.

\section{The shape of the DTC and $\kappa(t)$ in a linear system}
\label{sec:shape}

Since the DTC captures crucial properties related to spike time distributions, we next set out to explore possible relationships between the dynamical properties of the underlying dynamical system and the shape of the resulting DTC. We explore these questions considering a family of linear planar models with thresholds. While these provide excessively simplistic representations of neural dynamics and in particular all show type 3 excitability, their simplicity also enables full control of some key dynamical properties of the system, thus opening the way to a principled exploration of possible relationships between DTC and properties of the dynamics.  

These models are given by the equations 
\begin{equation}\label{eq:2dlinear}
\begin{cases}
dv=(av+bw+I(t))dt+\sigma dW_t\\
dw=(cv+dw)dt,
\end{cases}
\end{equation}
with coefficients $\{ a, b, c, d \}$, periodic input $I(t)$ given by the tent case \jr{(\ref{eq:tent})}, and noise of strength $\sigma >0$ that is included in the same way as in the $V-U$ model (\ref{eq:VU}).
We define model spikes by increases of $v$ through a constant threshold $\theta$ and impose a resetting mechanism on both variables:

\begin{equation}\label{eq:resetting2}
v(t^{*-})=\theta\  \implies \ v(t^{*+})=v_r, \ w(t^{*+})=w_r.
\end{equation}
In our simulations, we will take $v_r=w_r=0$, whereas the threshold $\theta$ will be set in a case-specific way, to appropriately modulate the minimal distance of the deterministic trajectory to the spiking threshold and ensure comparable experiments across model parameter sets.  

Let us denote by $M$ the matrix
\begin{equation}\label{eq:M}
M:=\begin{bmatrix}
a &  b\\
c & d
\end{bmatrix},
\end{equation}
the eigenvalues of which determine the stability type and the rate of attraction or repulsion associated with the unique equilibrium $(0,0)$ of the autonomous, deterministic ($\sigma=I(t)=0$) version of system \eqref{eq:2dlinear}. We will consider parameter values for which $(0,0)$ is an attracting focus. 
All models in this class are type 3: whatever the value of a constant input $I$, the system features a single fixed point, whose stability is just determined by the eigenvalues of the matrix $M$. This class of models thus allows us to explore any possible effects of the strength of the contraction and oscillation frequency of trajectories to the fixed point, determined by the real and imaginary parts of the eigenvalues of $M$, on response precision.  We shall see, however, that no simple relationship emerges between these parameters and the properties of the DTC.

Changing the coefficients of the matrix $M$ may lead to drastic changes in the shape of the trajectory of (\ref{eq:2dlinear}). The fluctuation process $\delta$ is also affected by the properties of the matrix $M$. In detail, $\delta$ is the planar Ornstein-Uhlenbeck process given by:
\[d\delta(t)=M\delta(t)\,dt + \binom{\sigma }{0}dW_t,\]
the stationary solution of which is well-known~\cite{risken} to be a Gaussian process with covariance matrix $\Sigma$ given by the solution of the Lyapunov equation
\[M\Sigma +\Sigma M^T = \left(\begin{array}{cc} \sigma^2 & 0\\
0 & 0 
\end{array}\right).\]
This solution can be calculated explicitly as:
\[\Sigma = \nu^2 \left(\begin{array}{cc} \sqrt{1+\frac{d^2}{\Delta}} & \frac{cd}{\sqrt{\Delta}} \\
\frac{cd}{\sqrt{\Delta}}  & \frac{c}{\sqrt{\Delta}} 
\end{array}\right),\]
 with $\nu:=\frac{\sigma}{\sqrt{2\vert \tau\vert}}$ where $\tau$ is the trace of $M$ and $\Delta$ is its determinant. In particular, we observe that the coefficient $\nu$ captures the order of magnitude of the fluctuations of the process resulting from the competition of the noise strength $\sigma$ and the linear contraction rate $\vert \tau\vert$. To     balance this effect and explore the role of the real part of the eigenvalues irrespective of its impact on the amplitude of the fluctuations, we adjusted the threshold to be placed at a distance from the trajectory proportional to $\nu$. Specifically, for each matrix $M$ with trace $\tau$, let $\max(v_{\sigma=0})$ be the supremum of the $v$-variable for the linear, deterministic ($\sigma=0$) case of model (\ref{eq:2dlinear}) starting from initial condition $(0,0)$, with no reset condition. The threshold for stochastic simulation is then set as 
\begin{equation}\label{eq:defTheta}
\theta:=\max(v_{\sigma=0})+\alpha \nu,
\end{equation}
where $\alpha$ is some positive constant. With that choice, keeping $\alpha$ fixed across various conditions ensures that
the minimal distance between the trajectory and the threshold remains the same in units of fluctuation amplitude. In particular, a change in the shape of the deterministic trajectories will be accounted for by translating the thresholds accordingly; for example,   the potential facilitation of spike emission due to an increase in the noise strength $\sigma$ is compensated for by the increase in the distance to threshold, and the potential hindering of spike emission due to an increased contraction is compensated for by a decrease in threshold. This setup  assures that the results of our numerical experiments stem from the dynamical properties of the system captured in matrix $M$.


 
For a linear system (\ref{eq:2dlinear}) with an asymptotically stable focus at $(0,0)$, we expect that the real parts of the eigenvalues of $M$, which control the strength of contraction of the flow towards the origin, the imaginary parts of the eigenvalues of $M$, which control the frequency with which trajectories circulate around the origin, and the 
nullcline slopes, 
which control the orientation of trajectories relative to the coordinate axes and hence relative to the $\{ v = \theta \}$ threshold, should each influence the DTC.  We therefore performed experiments in which we systematically varied one of these three factors, while holding the other two constant. Our methodology relies on basic linear algebra and is detailed in Appendix~\ref{sec:Matrices}.
%
 \begin{figure}[h!]
\begin{centering}
\hspace{-0.6in}
\includegraphics[width=\textwidth]{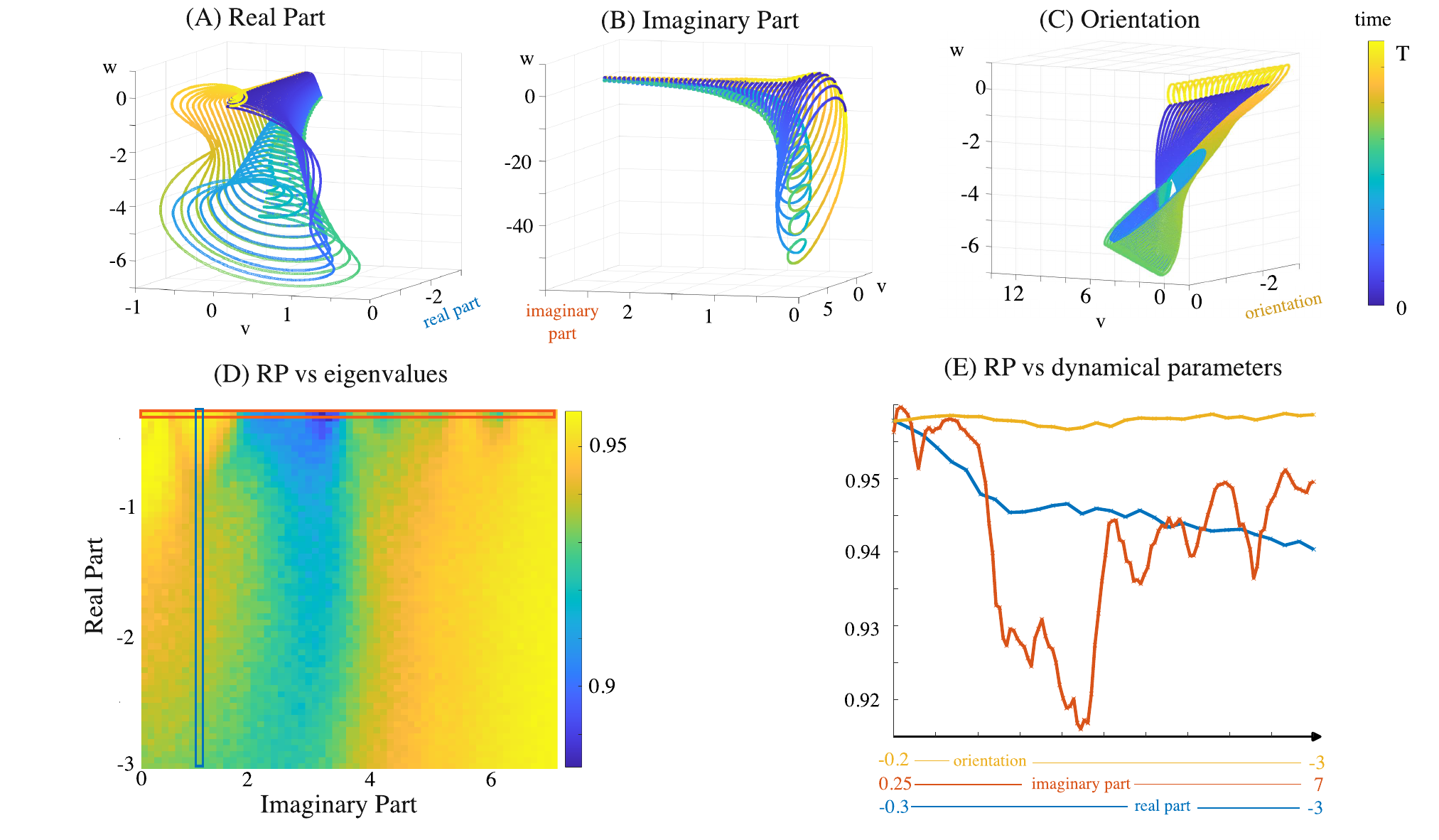}
\caption{The relation between the coefficient matrix $M$ and phase-locking is non-trivial for a linear system. Families of deterministic trajectories, color coded by time within one input period (shared color bar), as (A) the real part of the eigenvalues of $M$, (B) the magnitude of the imaginary part of the eigenvalues of $M$, or (C) the $w$-nullcline slope is systematically varied. Note the use of different viewing angles in different panels, to highlight the structure present in each. (D) Heat map of RP as eigenvalues' real and imaginary parts are varied. (E) Slices through the RP heat map from (D), with real part fixed and imaginary part varied over the indicated range (red horizontal band at the top of D, red curve in E), or vice versa (blue vertical band in D, blue curve in E), or with both held fixed as the $w$-nullcline slope is varied (orange). The baseline parameter values for these simulations were $a=-0.4, b=1.0, c=-1.5, d=-0.2$.} \label{fig:AUC_parameters}
\end{centering}
\end{figure}
Figure \ref{fig:AUC_parameters}A-C display families of deterministic ($\sigma=0$) trajectories   of (\ref{eq:2dlinear}) for one input cycle with initial condition (0,0) as one of the three matrix characteristics was varied. We observed that the amplitude of the trajectory decreased as the real part was made more negative, as expected with increased contraction, or when the imaginary part $\omega$ increased, with both effects analytically related to the amplitude of the solution of a linear differential equation with linear forcing,
\[\int_0^t s e^{(-\lambda +i\omega)s}\,ds = \frac{e^{-\lambda t}}{\omega^2} (-1+e^{i\omega t}(1-i\omega t)),\]
which decreases as $e^{-\lambda t} \omega^{-1}$.  As was also expected, we observed that the orientation of the trajectories twisted according to the orientation of the nullclines.


We systematically computed the RP associated with each of these trajectories, with a fixed noise level $\sigma$ and spike threshold scaled relative to both the noise level and the trace of $M$ according to~\eqref{eq:defTheta}. The resulting RPs associated with these changes in eigenvalues were complicated and difficult to predict or to correlate with any particular property of the trajectories.  Relative to the other features, the imaginary part of the eigenvalues had a prominent impact on RP, with a non-monotonic profile and reduced RP for imaginary parts in a certain interval of values (red outline in Figure~\ref{fig:AUC_parameters}D and red curve in Figure~\ref{fig:AUC_parameters}E). RP depended more weakly on the eigenvalues' real part, exhibiting an essentially monotonic, gradual decrease as the real part became more negative (Figure~\ref{fig:AUC_parameters}D-E, blue rectangle and blue curve). Lastly, despite its impact on the alignment of the trajectory relative to the threshold, the nullcline orientation unexpectedly did not  induce a noticeable effect on the RP
(Figure~\ref{fig:AUC_parameters}E, yellow curve). These observations further underline the non-trivial  relationship of the DTC to preferred spike phases, again highlighting that this structure  is not simply a readout of trajectory shape.  Moreover, they suggest that the DTC may not be linked in a simple way to properties of linearization or local flow features relative to the resting critical point in the nonlinear case, since this link is not present even in the fully linear setting.

\begin{figure}[h!]
\includegraphics[width=\textwidth]{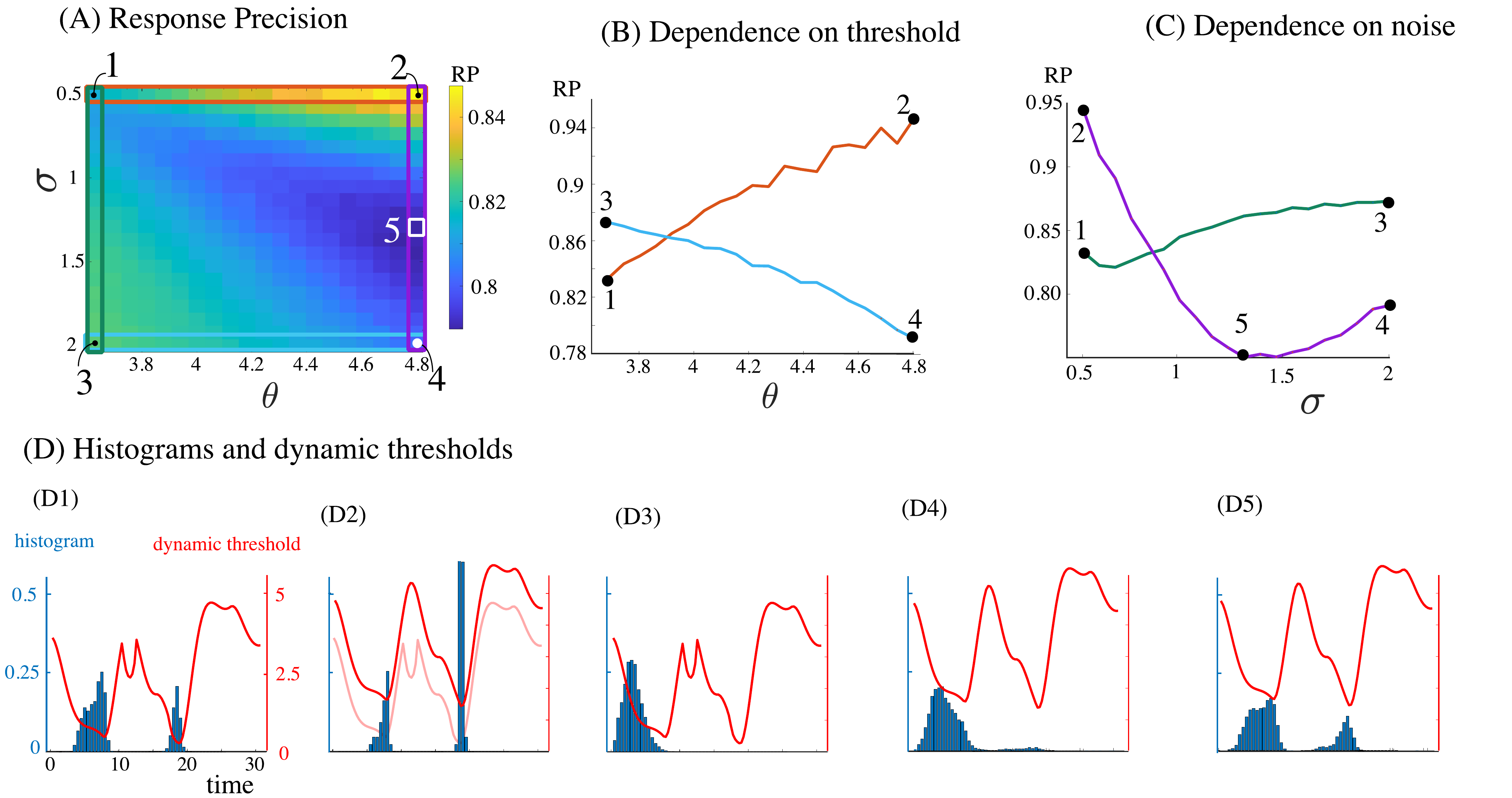}

\caption{ RP and spike distribution for the linear system (\ref{eq:2dlinear}) as a function of noise strength $\sigma$ and spike threshold $\theta$ as in (\ref{eq:resetting2}).  (A) RP (color) as a function of $\sigma$ and $\theta$. (B) RP  as a function of $\theta$ for two fixed values of $\sigma$ (red, small noise, and blue, large noise, corresponding to the horizontal matching color rectangles in (A)). (C)   RP  as a function of $\sigma$ for two fixed values of $\theta$ (green, small threshold, and purple, large threshold, corresponding to the vertical matching color rectangles in (A)). (D) Spike time histograms (blue bar plots) and dynamic threshold function (red curves), depicted on identical axes (common labels and axis values indicated on D1)  for five combinations of parameters labeled 1-5 (visible in (A) and identified on the curves of figures B and C). The pink curve in the graph in D2 corresponds to the dynamic threshold associated to point 1, and illustrates the change in dynamic threshold associated with an increase in spike threshold. All simulations here were run}  with $M=\left[ \begin{array}{cc} -0.4 & -0.68 \\ 1 & -0.8 \end{array} \right]$. \label{fig:linear_matrix1}
\end{figure}

Admittedly, enforcing an invariance of the distance to threshold (as measured in terms of a fraction of the fluctuation amplitude) might compensate for effects that would arise if the threshold, or noise strength, were kept constant. To round out our study of linear systems, Figure~\ref{fig:linear_matrix1} presents the RP and spike phase histograms of a linear system resulting from varying  $\sigma$ and the spike threshold $\theta$ independently.

Here again, RP exhibited complex, sometimes non-monotonic co-dependence on noise amplitude and spike threshold (Fig.~\ref{fig:linear_matrix1}A). In particular, we found that for small noise, increasing threshold led to greater RP, but for larger noise, increasing threshold had the opposite effect on RP (Fig.~\ref{fig:linear_matrix1}B). Increasing noise led to non-monotonic RPs (Figure~\ref{fig:linear_matrix1}C), with an overall increase in RP for small thresholds (green) and a decrease for large thresholds (purple). To appreciate more finely the phenomena leading to these complex dependencies, we focused on four specific situations (labeled 1 to 4 and corresponding to the corners of the heatmap in Fig.~\ref{fig:linear_matrix1}A, with spike time histogram and dynamic threshold function for each case shown in Fig.~\ref{fig:linear_matrix1}D). Corners 1 and 2 correspond to an identical, small noise level but distinct thresholds. Increasing threshold with small noise led to an overall increase in the dynamical threshold function (Fig.~\ref{fig:linear_matrix1}D2, comparing the red curve associated to Corner 2 and faint pink curve for Corner 1). In both cases, the spike histograms are bimodal, with peaks associated to each trough of the dynamic threshold function, but with larger $\theta$, both peaks are much narrower and more spikes occur at the second trough, which is also the lowest part of the dynamical threshold function.  Indeed, in the case of smaller $\theta$, with a lower dynamic threshold function, there is an increased probability that each cycle will cross through the dynamic threshold function at the first trough; moreover, a reduction in $\theta$ widens the range of phases at which the dynamic threshold is in within reach of typical fluctuations, which broadens the spike phase distribution.   

Interestingly, for larger noise values (that is, comparing corners 3 and 4), we found the opposite effect. In fact, for large levels of noise and small thresholds,  it becomes unlikely for trajectories to avoid crossing the leading slope of the first trough of the dynamic threshold function. The resulting spike distribution is thus unimodal and relatively concentrated. Increasing the threshold now relaxes this phenomenon, and by making the probability to spike very early smaller, it broadens the spike time distribution around the first trough of the dynamic threshold and also introduces the opportunity for trajectories not to spike until the second trough, leading to a decrease in RP. 

These interpretations also provide more insight about the dependence of RP on noise levels. When considering a fixed spike threshold $\theta$, the dynamic threshold is also fixed, so changes in noise amplitude affect the RP by altering the distribution of first passage times to a fixed curve. When this curve is low (comparing Corners 1 and 3), increasing noise will lead to a concentration of the spike distribution as trajectories are made more likely to cross the threshold early. Instead, for larger thresholds (comparing corners 2 and 4), stochastic trajectories with small noise have overall a small probability to cross the dynamic threshold, and when they do spikes are concentrated at its troughs, such that  increasing noise broadens the spike time distribution. 

Overall, both the interpretations that we have provided and the graph in Figure~\ref{fig:linear_matrix1}A show that there is a co-dependence of RP on noise strength and spike threshold, with contours of constant RP arising  along a diagonal axis.
Relatively high RPs can result from high spike thresholds with large enough noise to induce dynamic threshold crossings at the deepest troughs (Figure ~\ref{fig:linear_matrix1}D2) or from low thresholds with enough noise to ensure that all spikes occur in the early part of the first dynamic threshold trough (Figure ~\ref{fig:linear_matrix1}D3).  Especially low RPs occur in between these two regimes, where spikes are neither concentrated in the trough minima nor at the leading edge of the first trough and instead form a bimodal distribution with two broad peaks (Figure ~\ref{fig:linear_matrix1}D5).

\section{Dynamic thresholds and phase-locking in an auditory neuron model}
\label{sec:homotopy}

Our analysis of spike generation in terms of a first passage of a stochastic process to the DTC $\Phi(t)$ (or equivalently of a related fluctuation process to $\kappa(t)$) highlighted the fact that the shape of the DTC, together with noise levels, determines the spike phase distribution. Our investigation of two-dimensional linear systems showed that simple properties based on linearization of the underlying dynamical system fail to predict response precision  and hence further emphasized the importance of the DTC, which reflects the input structure, and the noise intensity.  To complete our analysis, we  return to the auditory neuron setting of the $V-U$ model and numerically investigate the shape of $\kappa(t)$ and the RP that it yields, under systematic variation of the model structure between Type 2 and Type 3 limits. 

Specifically, we introduce a generalized version of the $V-U$ model, of the form

\begin{equation}
    \label{eq:VUhom}
    \begin{array}{rcl}
    C_m dV & = & \displaystyle{\Big(I(t)-[\lambda g_{KLT}^{ton}+(1-\lambda)g_{KLT}^{pha}(U)](V-E_K)-} 
     \\
    & &\\
    & & \displaystyle{g_{Na}m_{\infty}^3(V)(aU/b)(V-E_{Na})-g_{KHT}(0.85n_0^2+0.15p_0)(V-E_K)}\\
    & &\\
    & & \displaystyle{-g_{L}(V-E_L)-g_h r_0 (V-E_h)\Big)\,dt + \sigma dW_t}  \vspace{0.1in} \\
    & & \\
    dU & = & \displaystyle{\frac{U_{\infty}(V)-U}{\tau_U(V)}\,dt}
    \end{array}
\end{equation}
where, as in Section \ref{sec:phaselock}, $g_{KLT}^{pha}(U)= \bar{g}_{KLT}a^4(1-U)^4z_0$ and
$ g_{KLT}^{ton}= \bar{g}_{KLT}a^4(1-U_0)^4z_0$ respectively
denote the forms of $g_{KLT}$ originally used in the phasic, or Type 3, and tonic, or Type 2, versions of the model \cite{meng2012}.  
In system (\ref{eq:VUhom}), $\lambda$ is a homotopy parameter, so that $\lambda=0$ corresponds to the Type 3 regime and $\lambda=1$ to the Type 2 regime.

\begin{figure}[h!]
\begin{centering}
\includegraphics[width=\textwidth]{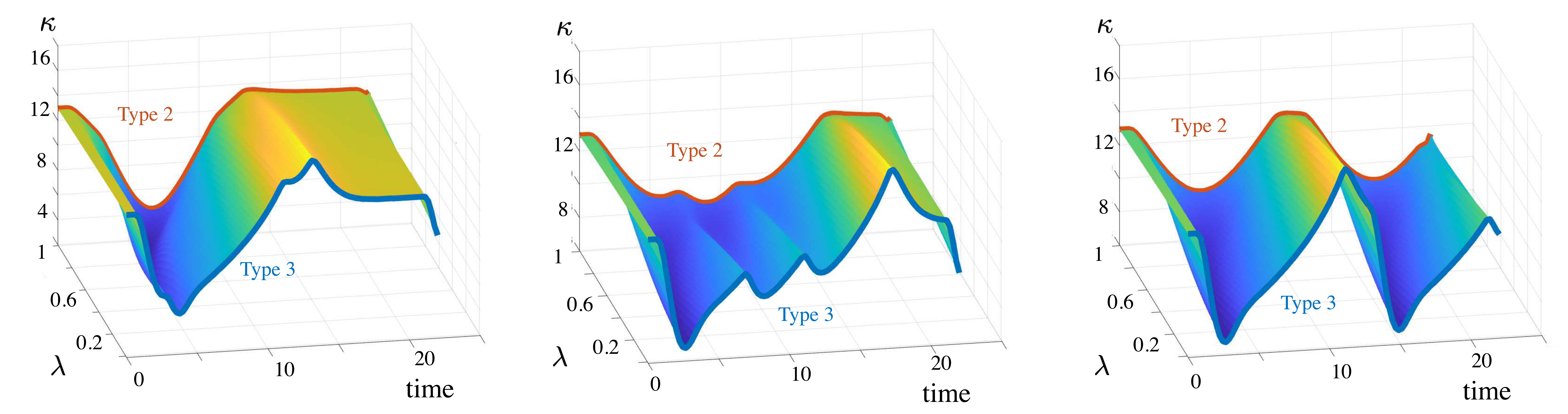}
\caption{Surface of $\kappa(t)$ over one input period as $\lambda$ (homotopy parameter) is varied for $\Delta T=2$ (left), $6$ (center) and $12$ (right). The extremes of $\lambda=1$ (Type 2 case) and $\lambda=0$ (Type 3 case) have their $\kappa(t)$  outlined in red and cyan, respectively.  The color-coding of the surface corresponds to height. }
\label{fig:homotopy}
\end{centering}
\end{figure}

Variation of $\lambda$ over the interval $[0,1]$ reveals a gradual but pronounced change in the shape of the curve $\kappa(t)$. Figure~\ref{fig:homotopy} depicts the shape of the dynamic threshold function as $\lambda$ is varied for three input profiles (superposition of 2 rectified sine waves for three values of the shift $\Delta T$ between those functions).  As $\Delta T$ increases, the emergence of a second input peak is reflected in the development of a second trough in the dynamic threshold function across all values of $\lambda$.  Notably, troughs in the Type 3 regime are consistently deeper and narrower than for Type 2. The middle of the three cases shown in Figure \ref{fig:homotopy} represents a particularly striking example, where the DTC structure in the Type 2 case includes what is effectively one wide trough with a mildly non-monotonic height whereas the Type 3 case features a clearly dominant, deep and narrow first trough, followed by two much shallower troughs.

\begin{figure}[h!]
\begin{centering}
\includegraphics[width=0.85\textwidth]{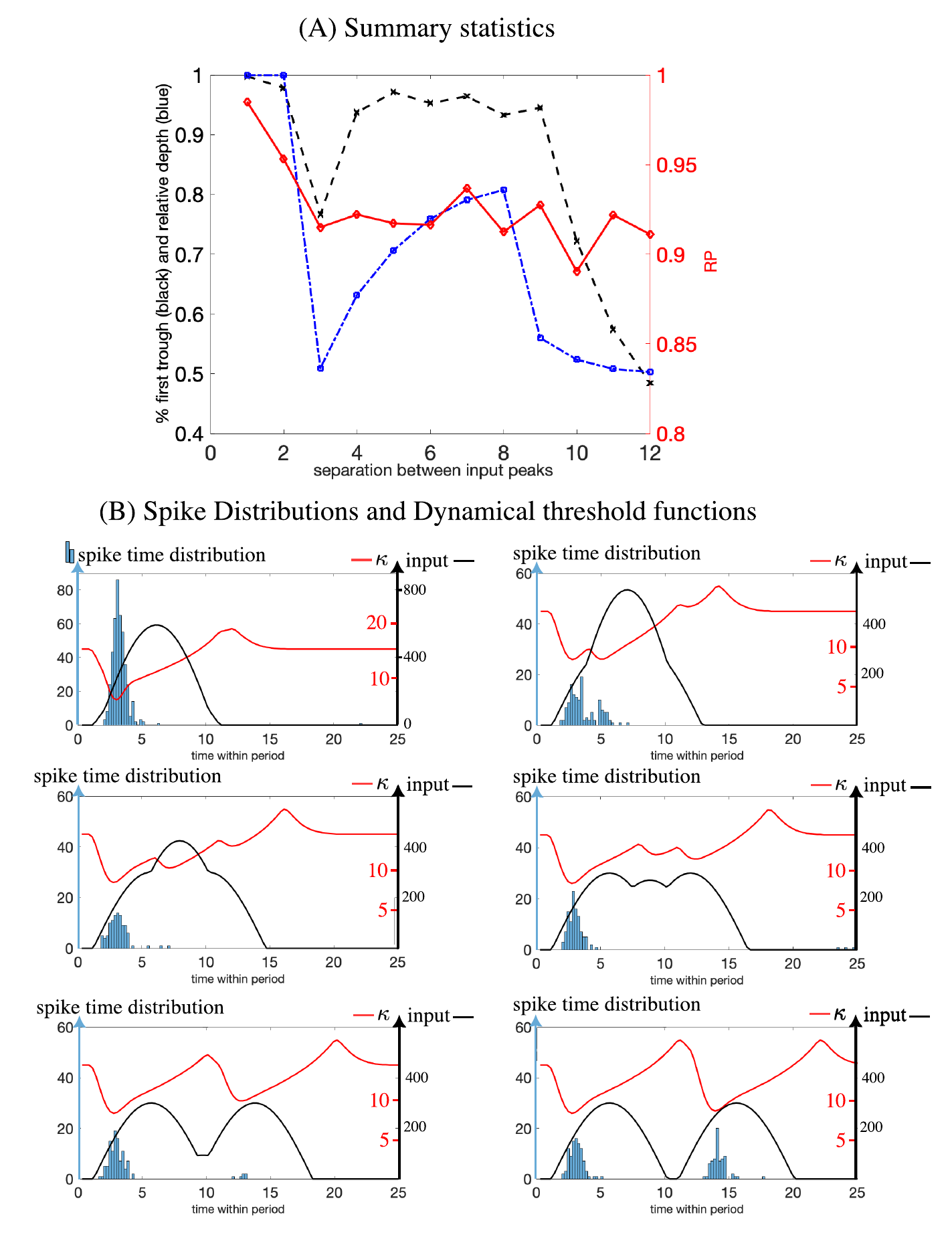}
\caption{Spike distribution and dynamic threshold function ($\kappa(t)$) properties as $\Delta T$ increases in the Type 3 case, run for 30000 time units with input amplitude 600 and noise amplitude $\sigma=7.0$.
(A) Percentage of spikes in the first spike histogram peak (black dashed), relative depth of the two troughs of $\kappa(t)$ (blue dash-dotted), and response precision (RP, red solid) versus $\Delta T$. (B) Spike statistics and dynamic threshold function for $\Delta T$ ranging in steps of 2 from 1 to 11. Blue histograms: spike phase distributions, red curves: dynamic threshold function $\kappa(t)$, black curves: input. }
\label{fig:phasicTE}
\end{centering}
\end{figure}

\begin{figure}[h!]
\begin{centering}
\includegraphics[width=0.85\textwidth]{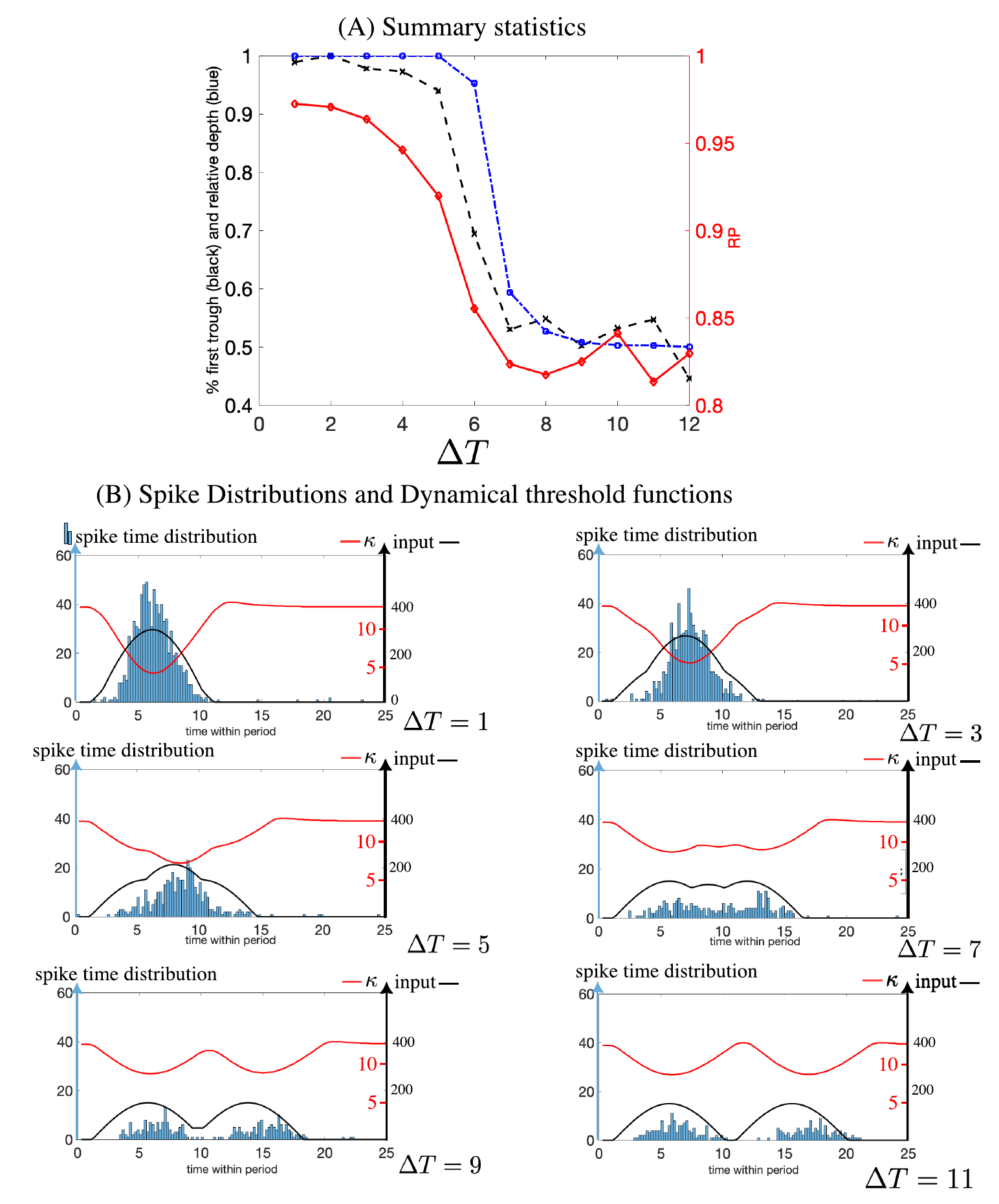}
\caption{{Spike distribution and dynamic threshold $\kappa(t)$ properties as $\Delta T$ increases in the Type 2 case, with $g_{KLT}^{ton}=9.17$, $g_{Na}=1500$, input amplitude 100 and noise amplitude $\sigma=6.0$. Same layout and color code as in Figure~\ref{fig:phasicTE}. }
}
\label{fig:tonicTE}
\end{centering}
\end{figure}

We next investigated how the trough structure emerges as $\Delta T$ increases and how this impacts the distribution of spike phases in the two models.
Specifically, over a sequence of increases in $\Delta T$ from 1 to 12, in each of the Type 3 and Type 2 models, we quantified the percentage of spikes in the first peak of the spike distribution histogram (corresponding to the first trough of the graph of the dynamic threshold function $\kappa(t)$), the relative depths of the troughs, and the RP (see Appendix \ref{sec:PFP_DTC_nAUC_numerics} for methodological details); see Figures \ref{fig:phasicTE}, \ref{fig:tonicTE}.  In the Type 3 regime (Figure \ref{fig:phasicTE}), as $\Delta T$ is increased, the additional structure of the input pattern translates into the formation of secondary troughs in $\kappa(t)$ that can be viewed as non-monotonicities on the back side of the leftward-skewed primary trough. As secondary troughs are first spawned, the mode of the spike histogram splits across them, causing a drop in the percentage of spikes in the first peak (PFP) and in the RP.  With larger $\Delta T$, the smaller troughs pull towards higher phases and the first spike peak in the histogram is abruptly restored, leading to a sharp recovery of the PFP.  Gradually, these troughs merge into a significant secondary trough, after which the PFP drops sharply.  Although the secondary trough yields a secondary mode in the spike histogram, the RP does not drop much, because each trough is sharp and each of the two peaks remains concentrated over a narrow range of phases.

In the Type 2 case (Figure \ref{fig:tonicTE}), the situation is qualitatively different and simpler, with essentially monotone changes in $\kappa(t)$ trough and spike histogram properties as $\Delta T$ grows.  Specifically, an increase in $\Delta T$ results not in a trough splitting but rather in a symmetric perturbation to the trough shape; that is, the single trough itself remains at least roughly symmetric relative to its center.  The spike histogram remains unimodal throughout these changes yet broadens, resulting in a gradual decline in the RP.  Eventually, a local maximum arises within the $\kappa(t)$ trough, which splits it into two parts of roughly equal depth.  By that time, the spike distribution is quite wide, and it becomes split into two roughly equal parts, with a saturation of the RP at a relatively low plateau, reflecting the inferior phase-locking in this regime.

\begin{figure}[h!]
\begin{centering}
\includegraphics[width=\textwidth]{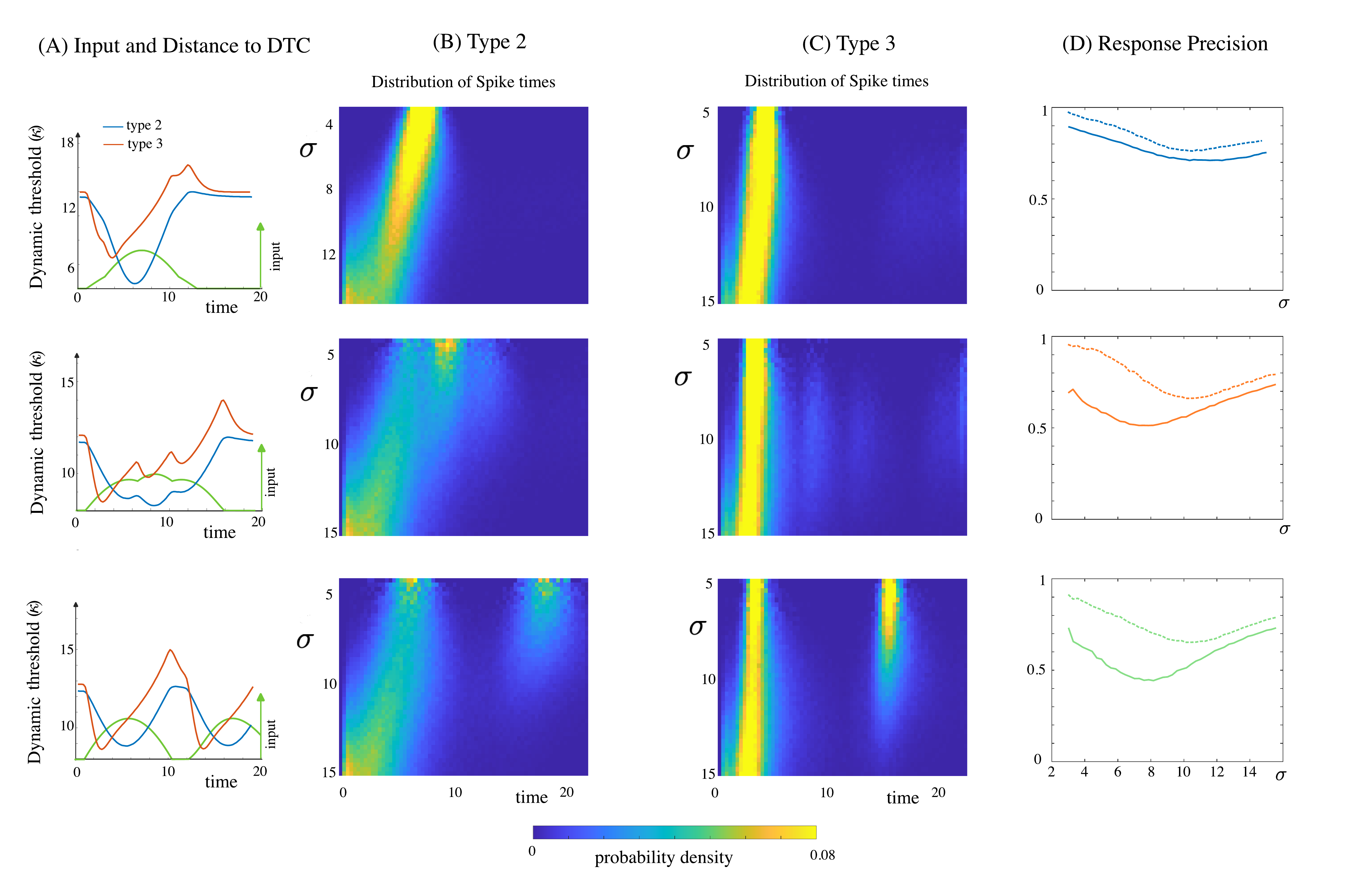}
\caption{Dynamic thresholds and phase locking in the Type 2 and Type 3 cases. 
Each column includes results for $\Delta T \in \{2,6,12\}$ (from top to bottom).  (A) Input (green) as a function of time within one cycle, along with the dynamic threshold functions for Type 2 ($\lambda=1$, blue) and Type 3 ($\lambda=0$, red) regimes. The blue and red curves are identical to those depicted in Figure~\ref{fig:homotopy}. (B-C) Spike time probability distribution in the Type 2 or Type 3 case, with associated RP (D), showing a non-monotonic dependence in noise levels. }
\label{fig:type2type3}
\end{centering}
\end{figure}

Since we found that noise intensity $\sigma$ can strongly impact response precision in the linear case (Figure \ref{fig:linear_matrix1}), we examined the spike phase distribution across the Type 2 and Type 3 extremes for three cases of $\Delta T$, over a range of values of $\sigma$ (Figure \ref{fig:type2type3}). 
In general, for low noise, the probability of spiking on any one input cycle is small, and this probability rises with $\sigma$.  We compute the spike probability distribution as a function of phase for all cases, however, to obtain results that can be compared across $\sigma$ values. Of note, only distributions associated with a spike probability larger than 1\% per cycle were kept (note, by consequence, that the minimal values of $\sigma$ associated with different panels in Figure \ref{fig:type2type3} are distinct). 

When the input is unimodal and $\kappa(t)$ has a single trough (Figure \ref{fig:type2type3}, first row, column (A)), this trough is narrower, with steeper sides, for Type 3 than for Type 2 (red versus blue, see also Fig.~\ref{fig:homotopy}).  Correspondingly, although the two cases exhibit relatively similarly narrow spike phase distributions for small $\sigma$, the distribution broadens noticeably more as $\sigma$ increases in the Type 2 case (column B vs. column C).  Moreover, the distribution peak in the Type 2 case shifts to earlier phases, which is a general property in these systems that is fully accounted for by our description of spiking as the first passage time of a stochastic process to a curved boundary (Section \ref{sec:stochastic} and Figure~\ref{fig:linear_matrix1} in the linear case):  as the variance of a random walk up to the DTC increases, it can hit the DTC across a broader range of phases, and the first hitting time therefore occurs earlier.  This effect is much less evident in the Type 3 case because the skewed trough shape favors hitting at early phases for all $\sigma$.

When $\Delta T$ increases to a moderate value and the input becomes non-monotonic, the previously-noted strong distinction between the Type 2 and Type 3 dynamic threshold functions emerges (Figure \ref{fig:type2type3}, middle row, (A)). Not only does the Type 3 case exhibit a much more clearly dominant first trough, but this trough shifts to earlier phases relative to the Type 2 case. Correspondingly, the spike phase distribution is much broader for the Type 2 than Type 3 case for this $\Delta T$, even for low noise, and the Type 3 distribution lies in a narrow range at relatively small phases (Figure \ref{fig:type2type3}, middle row, (B)-(C)).  The Type 3 DTC trough remains steep for this $\Delta T$, leading to a maintenance of a narrow spike phase distribution even with significant increases in noise.

Finally, when $\Delta T$ is large enough to yield a full separation between input peaks, the dynamic threshold functions for both extreme cases naturally both have two troughs (Figure \ref{fig:type2type3}, bottom row, (A)).  Nonetheless, the Type 3 first trough maintains its steeper curvature and shift to earlier phases, relative to the Type 2 case.  Thus, although the spike phase distributions in both cases are clearly bimodal,  the distribution is much broader for the Type 2 case  than for Type 3, with a gradual shift in the first peak location towards earlier phases as $\sigma$ increases, similarly to the other cases (Figure \ref{fig:type2type3}, bottom row, (B)-(C)).

\section{Discussion}
\label{sec:discuss}
To determine the possible dynamical structures controlling the precision of neural firing in response to periodic signals, we investigated in detail the phenomenon of phase-locking in non-autonomous excitable dynamical systems subject to stochastic fluctuations. While much is known about spontaneous dynamics or responses to pulses or steps of currents, the spiking activity of excitable systems driven by a time-dependent input, especially in the presence of noise, has received significantly less attention. The difficulty of the problem lies in the fact that whether the neuron will fire a spike or not depends not only on the current position of its trajectory but also on the future input it will receive. We reasoned here that, for a prescribed input shape, such spiking events are also predictable, but their occurrence depends on the phase at which a perturbation arrives. Building upon this observation, a key innovation of this work is our introduction of the dynamic threshold function $\kappa(t)$ and the corresponding dynamic threshold curve (DTC). The DTC represents the effective spike threshold: at each time $t$, it separates the phase plane into a quiescent region around the deterministic, non-spiking trajectory, whereby the forward orbit passing through a point in this region at time $t$ does not fire a spike after $t$ despite the ongoing input, and a spike region 
from which trajectories do yield spikes in the presence of ongoing input.
The derivation of this curve thus incorporates the effects of the underlying dynamics, whether linear or nonlinear, together with the impact of the time-varying input. Along a prescribed perturbation direction, this dynamical threshold curve lies at a distance $\kappa(t)$ from the sub-threshold, periodically forced, deterministic trajectory; the map $\kappa$ thus provides an estimate of the system's sensitivity to perturbations at any given point in time. The fluctuations in the function $\kappa$ orchestrate the firing times: troughs in $\kappa$ correspond to highly sensitive windows, where stochastic perturbations can easily kick the system outside of the quiescent region and induce a spike, while high spots or even increasing points in $\kappa$ correspond to times where the system will be unlikely to spike. 

To establish these connections, we have estimated the fluctuation process around the deterministic trajectory for a stochastically forced system and have shown that the distribution of input phases at which a system spikes can be captured by the passage of an elementary stochastic process, matching the variance of the fluctuation process, to $\kappa(t)$.  The variance of the stochastic process, including its sensitivity to input phase, will depend on the underlying dynamical system, and this effect together with differences in DTC shapes yield distinct precisions in the phases of responses in different systems. 
This investigation is reminiscent of recent results reporting how a gradual change of a parameter~\cite{ashwin2012} may lead a multistable dynamical system to tip between equilibria when a quasi-static approximation would fail to induce tipping. In particular, Slyman and Jones recently applied large deviation theory to analyze the most probable tipping path in a prototypical one-dimensional system with a monotone increasing input plus noise \cite{slyman2023}. Applying these techniques in our higher-dimensional setting with non-monotonic forcing constitutes a natural direction for future consideration. In particular, it would be worthwhile to investigate possible connections between the most likely paths and the trough of the dynamical threshold, or to analyze the large-deviation functionals associated with different neural models, in order to investigate how these can predict their distinct behaviors.

With this connection established, it remained for us to understand what properties of the underlying dynamical system control the shape of the DTC and of $\kappa$. Our exploration of various nonlinear and even linear systems revealed a complex picture. In particular, these properties are not well captured by the local structure associated with the stable equilibrium point that exists in the absence of forcing (Section \ref{sec:shape}). Noise, dynamics as well as past and future values of the input interact in subtle manners to facilitate or hinder spike events that, while well captured by the DTC and $\kappa$, are difficult to predict from an inspection of the underlying dynamical equation. Our analysis did reveal a seemingly paradoxical non-monotonic dependence of RP on noise levels (Figure \ref{fig:linear_matrix1}).  As expected, low noise levels lead to high RP with spikes concentrated at the lowest points of the dynamic threshold function. This precision will be degraded as noise levels are increased, with spikes occurring across a broader range of phases in a distribution  sculpted by the features of $\kappa$ beyond just its minima. However, we also found that with additional increases in noise, threshold crossings will become more likely earlier in each cycle,  which can increase RP again.
We also found that the sharpness and skew of troughs were important to consider along with their depths; indeed, just as a left-skewed trough in an escape boundary yields a narrower distribution of points where the boundary is first reached by a random walk, a left-skewed trough in $\kappa$ produces a higher RP.

Our original motivation for studying thresholds and response precision for excitable systems with time-dependent input was the observation of highly disparate degrees of phase-locking across two versions of an auditory neuron model and its reduced formulation \cite{meng2012,huguet2017}.  These two versions correspond to structural differences in the model equations that yield Type 3 dynamics, or only brief responses to sustained increases in input, in one case, versus Type 2 dynamics, or sustained responses to sustained increases in input, in the other.  
In Type 2 neural models, excitability can be lost as a model parameter is varied, with the real part of an eigenvalue of an equilibrium point going to 0.  Our analysis shows that this change in linear structure does not directly explain the model's phase-locking properties.
This difference is, however, perfectly captured by the dynamic threshold functions $\kappa$, which have broader, shallower troughs in the Type 2 case than in the Type 3 regime (Section \ref{sec:homotopy}), perhaps related to effects in the underlying dynamics that pull trajectories more strongly to negative voltages in between input peaks in the Type 3 case. Our results on dynamic thresholds shed new light on the dynamical mechanisms underlying the previous observations about phase-locking. Moreover, in the Type 2 version of the model, the low-threshold potassium current $I_{KLT}$ has its gating variable pinned to a constant value, whereas this variable is dynamic in the Type 3 case.  Thus, this work adds to the growing body of literature highlighting the importance of how timescales are represented within neural models for shaping their activity patterns \cite{franci2018,john2023}.  Finally, past work has shown that dynamical properties such as a slope detection that had been identified in Type 3 models can be recast in terms of more precise conditions, which clarify that these properties are not restricted to Type 3 models alone \cite{rubin2021}.  Similarly, the results here show that rather than a model's Type 3 or Type 2 status, the shape of $\kappa(t)$ for the model is the arbiter of the phase locking that it exhibits, and this shape can differ even within a fixed model, depending on the input pattern and the nature of the actual spike threshold.

Our study leaves open many directions for future investigation.  One subtlety is that in our simulations with the auditory neuron model, to enhance computational efficiency, we recorded spike phases across many single-cycle input  presentations.  This contrasts with the linear case, where we reset the trajectory but not the input after each spike. We found that this methodological distinction did not affect our results on phase-locking and the strong correspondence between $\kappa(t)$ and the spike phase distribution, likely because in our case, $I(t)=0$ over a significant time interval at the end of each input period, which allowed ample time for trajectories to contract to a small neighborhood of the critical point before the next wave of input arrived.  With input patterns that do not feature this effective pause between cycles, phase-locking could weaken, and the DTC itself could become harder to define, because effective distance from threshold could depend on the number of cycles since a spike last occurred.  This complication is reminiscent of efforts to define phase response curves, which typically involve an assumption of rapid return to an underlying, unperturbed oscillation, in settings where this return is slower and may extend across multiple cycles \cite{oprisan2001,oprisan2004}. In terms of coding of input, the question of phase precision should further be explored in conjunction with the number of input cycles skipped. Indeed, while high phase precision could for instance be achieved with low levels of noise in certain conditions, low noise may also lead to low spiking probabilities, requiring many cycles, or the collective responses of large ensemble of cells, to provide accurate information to downstream areas, which could be problematic for inducing effective responses to stimuli.

Another subtlety is that while noise can push a trajectory across the DTC, it could also theoretically prevent spikes even after threshold is crossed.  We ignored this possibility for two reasons.  First, our simulations showed that after most perturbation-induced DTC crossings, the deterministic effects from model terms and input forcing acting on trajectories swept them rapidly from DTC to the spike threshold, easily dominating noise effects.  Second, we focused on the distribution of phases at which spikes occurred. We reasoned that the low-probability event of noise pushing a trajectory back from above to below the DTC would be roughly equally likely across all cycle phases, and hence this factor would have little influence on the overall spike distribution.  A related issue is that in theory there may be non-monotonicity in which a small increase in a variable induces a large response whereas a larger increase prevents it again, which would complicate even defining the DTC.

Our study introduces the DTC, establishes its utility, highlights its relation to phase-locking, and provides some theoretical insights based on a stochastic process perspective.  However, our investigations did not result in a detailed explanation of how features of an underlying dynamical system translate into specific $\kappa(t)$ and DTC properties.  Establishing such a deeper mechanistic understanding represents an important direction for future analysis. Moreover, although our DTC definition (Section \ref{sec:DTC}) allows for arbitrary numbers of dimensions, analogously to the situation for non-dynamic thresholds based on rivers \cite{letson2020}, we restricted our numerics to 2D dynamics.  Numerical estimation, visualization, and analysis of $\kappa(t)$ and the DTC in higher dimensions would involve a variety of challenges that we did not tackle and hence remain for future work.  
\newpage
\appendix

\section{$V-U$ model definition}
\label{sec:VUFunctions}
The $V$-dependent functions in system (\ref{eq:VU}) are given by the following:
\begin{equation}
    \label{eq:Vfunc}
    \begin{array}{rcl}
    m_{\infty}(V) & = & (1+\exp(-(V+38)/7))^{-1}, \vspace{0.1in} \\
    U_{\infty}(V) & = & b[h_{\infty}(V)+b(a-w_{\infty}(V))]/a(1+b^2), \vspace{0.1in} \\
    h_{\infty}(V) & = & (1+\exp((V-\theta_h)/\sigma_h))^{-1}, \vspace{0.1in} \\
    w_{\infty}(V) & = & (1+\exp(-(V-\theta_w)/\sigma_w))^{-1/4}, \vspace{0.1in} \\
    \tau_U(V) & = & \min \{ \tau_h(V),\tau_w(V) \}/3, \vspace{0.1in} \\
    \tau_h(V) & = & 1.5 + 100/[6\exp((V+60)/6)+16\exp(-(V+60)/45)], \vspace{0.1in} \\
    \tau_w(V) & = & 0.6 + 100/[7\exp((V+60)/11)+10\exp(-(V+60)/25)].  
    \end{array}
\end{equation}
The parameters $\theta_h, \sigma_h, \theta_w, \sigma_w$ take different values for the phasic model, namely $(-65,6,-48,6)$, and the tonic model, where they are $(-66,7,-46,7)$.
\begin{table*}
\caption{\label{tab:VUparam} Default parameter values for the $V-U$ model (\ref{eq:VU}),(\ref{eq:Vfunc}).}
\begin{center}
\begin{tabular}{|cr|cr|cr|cr|}
\hline
$g_{Na}$ & $1000.0$ nS & $g_{KHT}$ & $150.0$ nS & $g_L$ & $2.0$ nS & $g_h$ & $20.0$ nS \\ \hline
$E_{Na}$ & $55$ mV & $E_K$ & $-70$ mV & $E_L$  & $-65$ mV & $E_h$ & $-43$ mV  \\ \hline
$C_m$ & $6.0$ pF & $a$ & 0.9 & $b$ & 0.8742 & &  \\ \hline
$n_0$ & 0.0077 & $p_0$ & 0.0011 & $r_0$ & 0.147 & $z_0$ & 0.662  \\ \hline
\end{tabular}
\end{center}
\end{table*}

\section{Numerical methods}
\subsection{Dichotomic search for the DTC}\label{sec:numerics}
To compute the distance to threshold $\kappa$ and the DTC efficiently, we used a dichotomic search. This method consists of the computation of a sequence of nested intervals $[k_1^n,k_2^n]$ that enclose the value of $\kappa(t^*)$ at a given time $t^*$. Here we describe this method in more detail in the context of the two-dimensional problem outlined in Section~\ref{sec:FES}. The sequence of intervals is initialized by setting $k_1^0=0$ and $k_2^0$ to be some value such that $v(t)+k_2^0 p_1>v_{th}$. At each step, we compute the trajectory associated with a perturbation with amplitude $k_{mid}^n=\frac{k_1^n+k_2^n}{2}$. 
 \begin{itemize}
     \item If the trajectory computed does not cross the threshold before the time horizon $T$, when we set $k_1^{n+1}=k_{mid}^n$ and $k_2^{n+1}=k_2^n$. 
     \item If the trajectory computed crosses the threshold, when we set $k_1^{n+1}=k_1^n$ and $k_2^{n+1}=k_{mid}^n$. 
 \end{itemize} 
 As usual for dichotomic searches, the size of the interval is halved at every step, yielding high precision estimates in a relatively small number of steps.
 
\subsection{Numerical methods for Figures \ref{fig:phasicTE}, \ref{fig:tonicTE}}
\label{sec:PFP_DTC_nAUC_numerics}
To compute the percentage of spikes in the first peak (PFP) and the relative depths of the troughs in the graph of the dynamic threshold function $\kappa(t)$, which are plotted over the times within one input period in these figures, we first identified all interior local maxima and minima in our numerically computed representation of $\kappa(t)$, using the built in MATLAB commands \texttt{islocalmax} and \texttt{islocalmin}.  If only a single local minimum was present, then the relative depth and PFP were both set to 1.  If at least two local minima were present, then we considered the height of the local maximum between them.  When this height was sufficiently large relative to the two local minima, then the graph of $\kappa$ was considered to have two (or more) troughs.  In this case, the PFP was computed based on the fraction of spikes in the histogram at phases lying within the first trough.  The relative depth of the first trough was computed as $(\kappa_{mx}-\kappa_{mn1})/((\kappa_{mx}-\kappa_{mn1})+(\kappa_{mx}-\kappa_{mn2}))$, where $\kappa_{mx}$ denotes $\kappa$ evaluated at its local maximum between the two troughs and $\kappa_{mni}$ denotes $\kappa$ evaluated at its $i$th local minimum, $i \in \{1,2\}$. Finally, to compute the RP, we used the definition, including equation (\ref{eq:RP}), from Section \ref{sec:DTC}.
\

\subsection{Methodology to adjust eigenvalues and nullcline orientation}\label{sec:Matrices}
In section~\ref{sec:shape}, we present how the DTC is modified in a linear planar integrate and fire model as a function of eigenvalues' real and imaginary parts and nullcline orientation, for the model:
\[\dot{x}=Mx+\binom{I(t)}{0}\]
where $x=\binom{v}{w}$ and
\begin{equation}
M:=\begin{bmatrix}
a &  b\\
c & d
\end{bmatrix}
\end{equation}
is an arbitrary matrix with complex eigenvalues $\lambda = \mu\pm i \omega$. We systematically varied the coefficients of the matrix to modify independently $\mu$, $\omega$, and the slope of the $w$-nullcline, $v=-d\,w/c$. To achieve this, we used the classical relationships between the trace and determinant of the matrix and the eigenvalues. Recalling that the trace of a $2\times 2$ matrix with complex eigenvalues is twice the real part of the eigenvalues and the determinant is their squared modulus, we proceeded as follows:
\begin{enumerate}
\item[(i)] To create a matrix with eigenvalues $\lambda' = \mu'\pm i \omega$ and $w$-nullcline conserving a slope coefficient $d/c$, we (a) modified the coefficient $a$ to $a'=a+2(\lambda'-\lambda)$, thus ensuring that the trace (which is twice the real part of the eigenvalues) is $2\lambda'$. The impact of this change on the imaginary part was compensated by modifying the coefficient $b$ accordingly to ensure that the determinant was $\mu'^2+\omega^2$, which was achieved by modifying $b$ to $b'=-[\mu'^2+\omega^2-a'd]/c$. 
\item[(ii)] To modify the imaginary part of the eigenvalues to $\omega'$ without altering their real part or the slope of the $w$-nullcline, we updated the parameter $b$ only (which does not alter the trace - thus the real part - nor the nullcline slope) to $b'=[ad-(\lambda^2+\omega'^2)]/c$. \item[(iii)] To explore the role of the orientation of the $w$-nullcline without changing the eigenvalues, we modified the diagonal coefficient $d$ to, say, $d'$, compensated the impact of this change on the real part by adjusting $a$ to $a'=a+(d-d')$, and left the imaginary part invariant by changing $b$ to $b'=[a'd'-(\lambda^2+\omega^2)]/c$.
\end{enumerate}
\bibliographystyle{siamplain}
\bibliography{PL}
\end{document}